\documentclass[twocolumn,prd,footinbib,aps,prl,floats,floatfix,amsmath,amssymb,longbibliography,secnumarab
ic]{revtex4-1} %
\usepackage[final]{graphicx}
\usepackage{hyperref}
\usepackage{amsmath}
\usepackage{bbm}
\usepackage{amsfonts}
\usepackage{amssymb}
\usepackage{latexsym}
\usepackage{graphicx}
\usepackage[english]{babel}
\usepackage{multirow}
\usepackage{float}
\usepackage{url}
\usepackage{slashed}
\usepackage{xcolor} 
\usepackage[utf8]{inputenc}

\newcommand{\be}{\begin{equation}}
\newcommand{\ee}{\end{equation}}
\newcommand{\ba}{\begin{array}}
\newcommand{\ea}{\end{array}}
\newcommand{\bea}{\begin{eqnarray}}
\newcommand{\eea}{\end{eqnarray}}
\newcommand{\sss}{\scriptscriptstyle}

\newcommand{\nn}{\nonumber}

\newcommand{\W}{{\sss W}}

\newcommand{\bg}{{\rm bg}}

\begin{document}

\title{Fluid equations for fast-moving electroweak bubble walls}

\author{Benoit Laurent}
\author{James M.\ Cline}
\affiliation{McGill University, Department of Physics, 3600 University St.,
Montr\'eal, QC H3A2T8 Canada}
\begin{abstract}
The cosmological electroweak phase transition can be strongly first order in extended particle physics models.  To accurately predict the speed and shape of the bubble walls during such a transition, Boltzmann equations for the CP-even fluid perturbations must be solved.
We point out that the equations usually adopted lead to unphysical behavior of the perturbations, for walls traveling close to or above the speed of sound in the plasma.  This is an artifact that can be overcome by more carefully truncating the full Boltzmann equation.  We present an improved set of fluid equations, suitable for studying the dynamics of both subsonic and
supersonic walls, of interest for gravitational wave production and electroweak baryogenesis.
\end{abstract}
\maketitle

\section{Introduction}
The electroweak phase transition in the early universe is known to be a smooth-crossover within the standard model (SM), given the measured value of the Higgs boson mass \cite{Kajantie:1995kf,Kajantie:1996mn}.  The addition of new particles coupling to the Higgs can turn it into a strongly first order phase transition, proceeding by the nucleation of bubbles of the true, electroweak symmetry breaking vacuum, in the initially symmetric
plasma.  This possibility has been widely studied because of its potential for providing electroweak baryogenesis (EWBG) \cite{Trodden:1998ym,Cline:2006ts,Morrissey:2012db}, and gravity waves that might be observable in the upcoming LISA experiment \cite{Caprini:2015zlo,Caprini:2019egz}.

An important parameter for the efficiency of baryon or gravitational wave production is the terminal speed $v$ of
the bubble walls, with baryogenesis generally favoring slower walls, while faster walls tend to produce stronger gravity wave signals.  To determine $v$ and other relevant properties of the bubble wall, within a given particle physics model, one must self-consistently solve for the 
perturbations to the fluid induced by the wall; these are needed to determine the frictional force acting on the wall, that brings it to a state of steady expansion.

In previous literature on this subject, quantitative study of 
fast-moving walls has been hampered by an apparent singularity of the fluid equations occurring at the sound speed $c_s = 1/\sqrt{3}$, that we will explicitly demonstrate below.  This makes a microscopic calculation of the friction in such cases problematic, motivating phenomenological estimates for the friction  \cite{Espinosa:2010hh,Konstandin:2014zta,Huber:2013,Megevand:2010}, or else leaving aside
supersonic walls altogether \cite{Kozaczuk:2015owa}. Complementary approaches have been used to study the ultrarelativistic limit \cite{Bodeker:2009qy,Bodeker:2017cim,Megevand:2013hwa};  in this work
we are primarily interested in velocities $v\gtrsim c_s$ rather than $v\cong 1$.   We argue that the apparent sound barrier is an artifact of a particular truncation of the Boltzmann equations for the fluid perturbations, and that sensible solutions exist for wall speeds up to $v = 1$ by making
a better choice.

A similar observation was recently made in ref.\ \cite{Cline:2020jre} in the context of the CP-odd fluid perturbations that are needed to compute the source terms for EWBG, but the analogous study for the CP-even perturbations, relevant to determining the bubble wall properties, has not been done.  It requires more work because the perturbation in the local temperature $\delta\tau = \delta T/T$ (not needed for the EWBG source terms) must now be included in the network.  The optimal way of doing this turns out to be somewhat subtle, as we will discuss.

We start by reviewing the standard approach in section  \ref{OFsect} and the pathology of the perturbations it predicts for supersonic walls.   We derive improved fluid
equations in section \ref{NFsect}, and in section \ref{Solsect} the solutions of the old and new formalisms are compared for 
a typical background wall profile, as a function of the wall velocity $v$.  These results are used in section \ref{fric-sect} to compute the predictions for the friction
term in the Higgs field equation of motion, that determines
the bubble wall shape and speed.
There we highlight the problems with the old approach and their absence in the new one.  Conclusions are
given in section \ref{conc}.  Formulas for the coefficients of the new fluid equations are presented in Appendix \ref{appA},
and the results of refined estimates for the collision terms
are explained in Appendix \ref{appB}.

\section{Old formalism (OF)} 
\label{OFsect}
We begin by recapitulating the method that has been used in previous literature for computing
the plasma perturbations \cite{Moore:1995si,Moore:1995ua,John:2000zq,Huber:2011aa,Konstandin:2014zta,Kozaczuk:2015owa,Dorsch:2018pat}. These are the deviations of the distribution function $f$
for a given particle away from its equilibrium form, that  have been parametrized as \cite{Joyce:1994zt,Moore:1995ua,Moore:1995si}
\bea
\label{dfeq}
f &=& {1\over e^X\pm 1} = {1\over e^{\beta\gamma(E-v p_z) -\delta X} \pm 1},\\
\delta X(z) &=& \mu+\beta\gamma[\delta\tau(E-v p_z)+u (p_z-vE)]\nn
\eea
where $\beta = 1/T$, $\gamma = 1/\sqrt{1-v^2}$ and the equilibrium 
part, with $\delta X = 0$, is expressed in the rest frame of the bubble wall, taken to be planar and moving to the left.  $\mu$ is the dimensionless chemical potential (in units of temperature) and $u$ is the velocity perturbation.  The wall frame is convenient for expressing the Boltzmann equation since the solutions in this frame are stationary,
\be
\label{boltzeq}
{\mathbf L}[f] = \left({p_z\over E}\partial_z - {(m^2)'\over 2E}\partial_{p_z}\right)(f_v + \delta\! f) \cong -{\cal C}[f] 
\ee
where $\delta\! f = -(df_v/dX)\,\delta X \equiv -f_v'\,\delta X$ is the perturbation, 
$(m^2)' = dm^2/dz$ for a particle whose mass depends on the background Higgs field $h(z)$ (and possibly other fields like a singlet scalar) in the wall, and $f_v$ is the equilibrium distribution in the wall frame.

To approximately solve eq.\ (\ref{boltzeq}), three moments are taken,
by integrating over momenta $\int d^{\,3}p$ with the respective
weight factors \footnote{In the fluid frame these are simply $1$, $p_z$ and $E$.} $1$, $\gamma(p_z-v E)$ and $\gamma(E-v p_z)$, giving three coupled ordinary differential equations for the perturbations $q\equiv (\mu,u,\delta\tau)^{\intercal}$, that can be written in the $3\times 3$ matrix form
\be
\label{matbe}
    A_v q' + \Gamma q = S
\ee
with a rate matrix $\Gamma$ from the moments of the collision
term ${\cal C}$ and a source $S \sim v \beta^2(m^2)'$ from the 
Liouville operator ${\mathbf L}$ in (\ref{boltzeq}) acting on $f_v$.

The $A_v$ matrix depends on $v$ in such a way that $A_v^{-1}\Gamma$ becomes singular at $v = c_s$, and has only positive eigenvalues 
for $v > c_s$.  By constructing a Green's function to solve eq.\ (\ref{matbe}) \footnote{Strictly speaking, this method only works when the $z$-dependence of $A_v^{-1}\Gamma$ can be ignored on either side of the wall, but the same conclusion is borne out by a full numerical solution.}, one can see that this implies that the perturbations $q$ must strictly vanish in front of the wall for $v> c_s$.  Ref.\ \cite{Cline:2020jre}
has argued that this kind of behavior is unphysical,
since the fluid equations (\ref{matbe}) describe particle diffusion, which is a physically distinct process from the propagation of sound waves.  There is no reason why diffusion should be suddenly quenched
in the vicinity of a supersonic wall, since some fraction of particles in front of the wall can still travel fast enough to get ahead of it.

\section{Improved fluid equations (NF)}
\label{NFsect}
In this section we propose a new formalism (NF) for the fluid equations, motivated by the recent paper \cite{Cline:2020jre}.  In that work, 
the problem of 
artificial suppression of diffusion for supersonic
walls was overcome, following a long-established method of dealing with the velocity perturbation $u$ \cite{Cline:2000nw,Fromme:2006cm,Fromme:2006wx}.
The adoption of a specific form for  $u$ is known to lead to unphysical results, that can be avoided by instead writing the perturbations in the form
\be
f = f_v - f_v'\,\delta \bar X + \delta\! f_u
\ee
where now $\delta \bar X$ omits the velocity perturbation $u$, which is instead encoded through $\delta f_u$ in such a way that 
\be
    u \propto  \int d^{\,3}p\, {p_z\over 
    E}\,\delta\! f_u
    \hbox{\ \ and \ } \int d^{\,3}p\, \delta\! f_u = 0\,.
\ee
To deal with other integrals involving $\delta f$, one makes a factorization ansatz
\be
\label{substitution}
    \int d^{\,3}p\, Q\,\delta\! f_u \to u \int d^{\,3} p\,
        Q {E\over p_z}\, f_v
\ee
for any quantity $Q$.  This procedure was shown in ref.\ \cite{Cline:2000nw} to lead to nonsingular diffusion in front of supersonic walls, so long as one
carefully evaluates the full $v$-dependence of $A_v$, rather than linearizing it in $v$, and weighting the
Boltzmann equation by the moments $1$, $p_z/E$.

However ref.\ \cite{Cline:2000nw} only considered the 
case of CP-odd perturbations, where $\delta\tau$ plays no significant role and hence was omitted.  Our purpose in this work is to extend those results to include $\delta\tau$, whose value is needed for the full solutions to the field equations determining the shape and speed of the bubble walls.  To determine this additional perturbation, a third moment is needed.  We find that by choosing the weighting factor $E$, in addition to $1$ and $p_z/E$ (all defined in the wall frame),
the resulting $A_v$ matrix  becomes
\be
\label{Amatrix}
    A_v = \left(\ba{ccl}  
    C_v^{1,1} & \gamma v C_0^{-1,0} & D_v^{0,0}  \\
    C_v^{0,1} & \gamma(C_v^{-1,1}-v C_v^{0,2}) & D_v^{-1,0} \\
    C_v^{2,2} & \gamma(C_v^{1,2}-v C_v^{2,3}) & D_v^{1,1}
    \ea\right)
\ee
where the dimensionless functions $C_v^{m,n}$ and $D_v^{m,n}$ are defined as
\bea
\label{Cmn}
    C_v^{m,n} &=& T^{m-n-3}\int\frac{d^3p}{(2\pi)^3}\frac{p_z^n}{E^m}(-f_v')\,, \\
    D_v^{m,n} &=& T^{m-n-3}\int\frac{d^3p}{(2\pi)^3}\frac{p_z^n}{E^m}f_v\,. \nn
\eea
With this choice, $\det A_v$ 
has no singularity for wall speeds between $v=0$ and $1$, and it gives the desired
behavior in which diffusion ahead of the wall only
gets suppressed in the limit $v\to 1$. The source term becomes
\be
\label{source}
    S =\gamma v\frac{(m^2)'}{2T^2}\left(\ba{c}
    C_v^{1,0} \\
    C_v^{0,0} \\
    C_v^{2,1} \\
    \ea\right)\,.
\ee

In previous literature, the coefficients corresponding to $C_v^{m,n}$ and $D_v^{m,n}$ were usually calculated in the limit of vanishing mass (as well as only leading order in $v$), but we find that the variation of $m^2(z)$  for the relevant particles within the wall can have a significant impact on the shape of the solutions. We thus retain the full mass- and $v$-dependence of those functions. Moreover, it is possible to analytically determine the $v$-dependence by boosting to the plasma frame (see Appendix \ref{appA}).

We have also updated the components of the collision matrix $\Gamma$ to account for the new choice of moments. The calculation of ref.\ \cite{Moore:1995si} is improved by correcting some errors pointed out in ref.\ \cite{Arnold:2000} and by using a Monte Carlo algorithm to compute more accurately the collision integrals. The new values of the collision terms are given in Appendix \ref{appB}.

\begin{figure*}[ht]
\centerline{$$\qquad v=0.2 \qquad\qquad\qquad\qquad\qquad v = 0.5\qquad\qquad\qquad\qquad\qquad\quad v=0.7\qquad\qquad\qquad\qquad\qquad\qquad v=0.95  $$}
\centerline{$\!\!\!\!\!$\raisebox{0.5cm}{\rotatebox{90}{Old Formalism}}\includegraphics[scale=0.3]{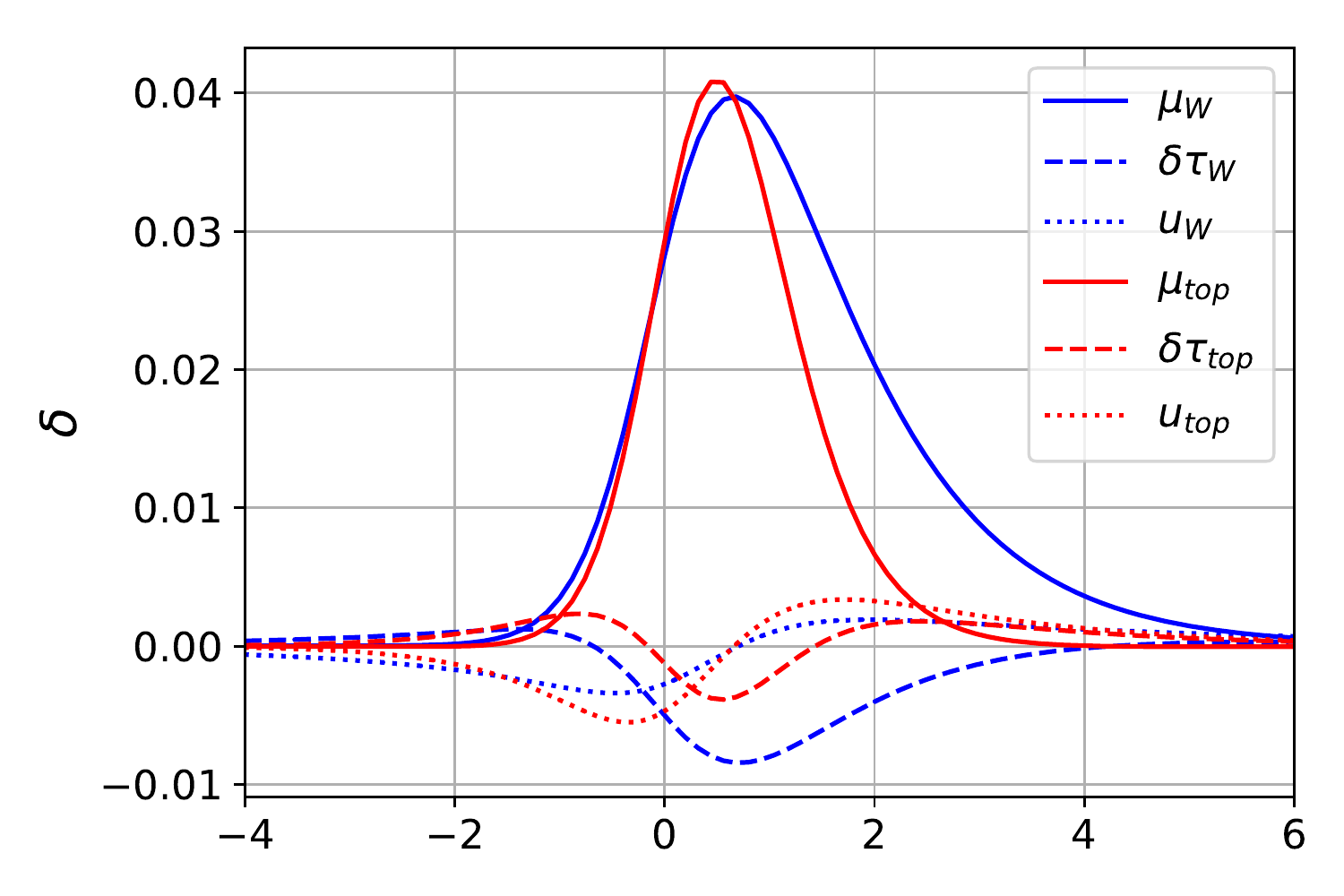}\includegraphics[scale=0.3]{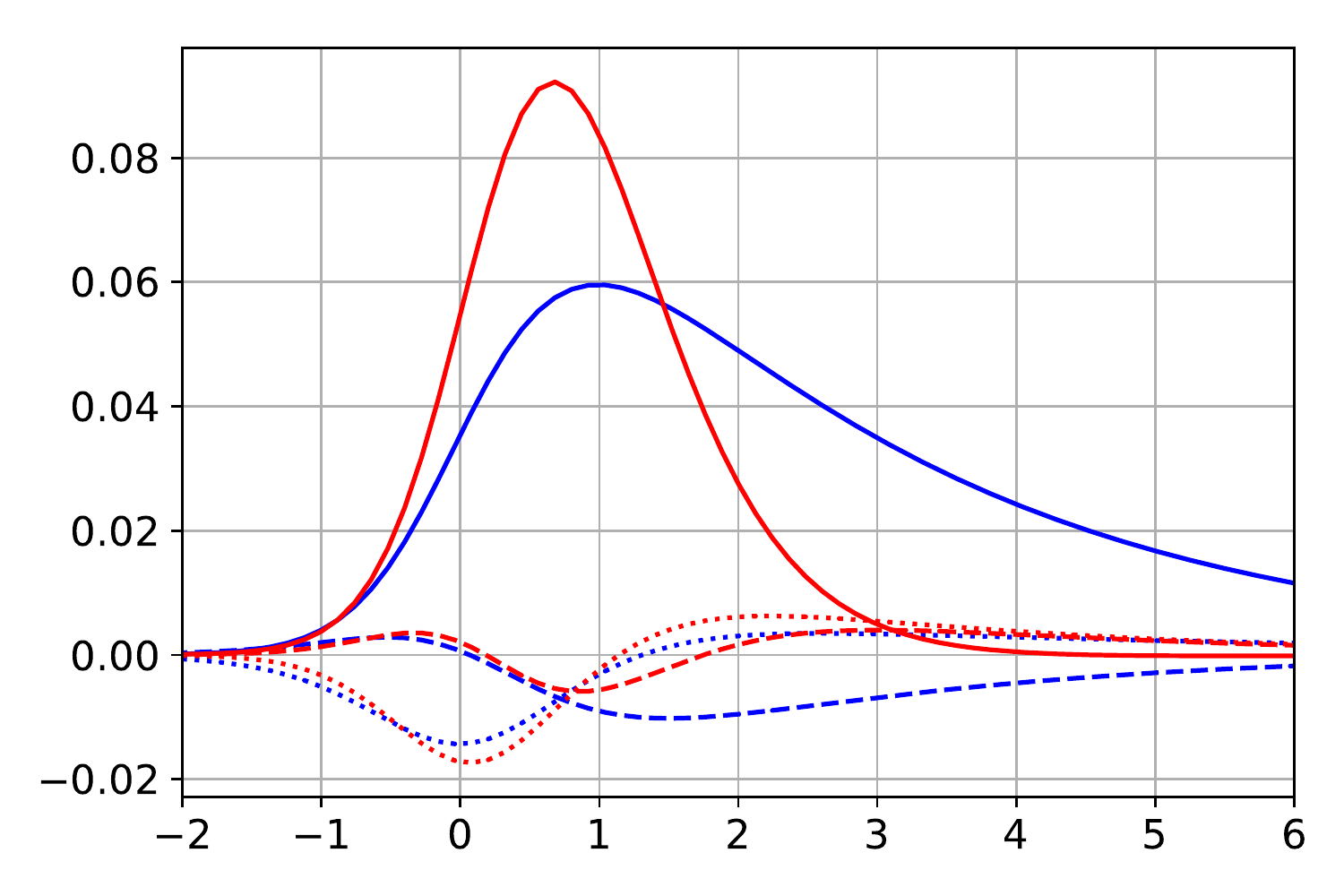}
\includegraphics[scale=0.3]{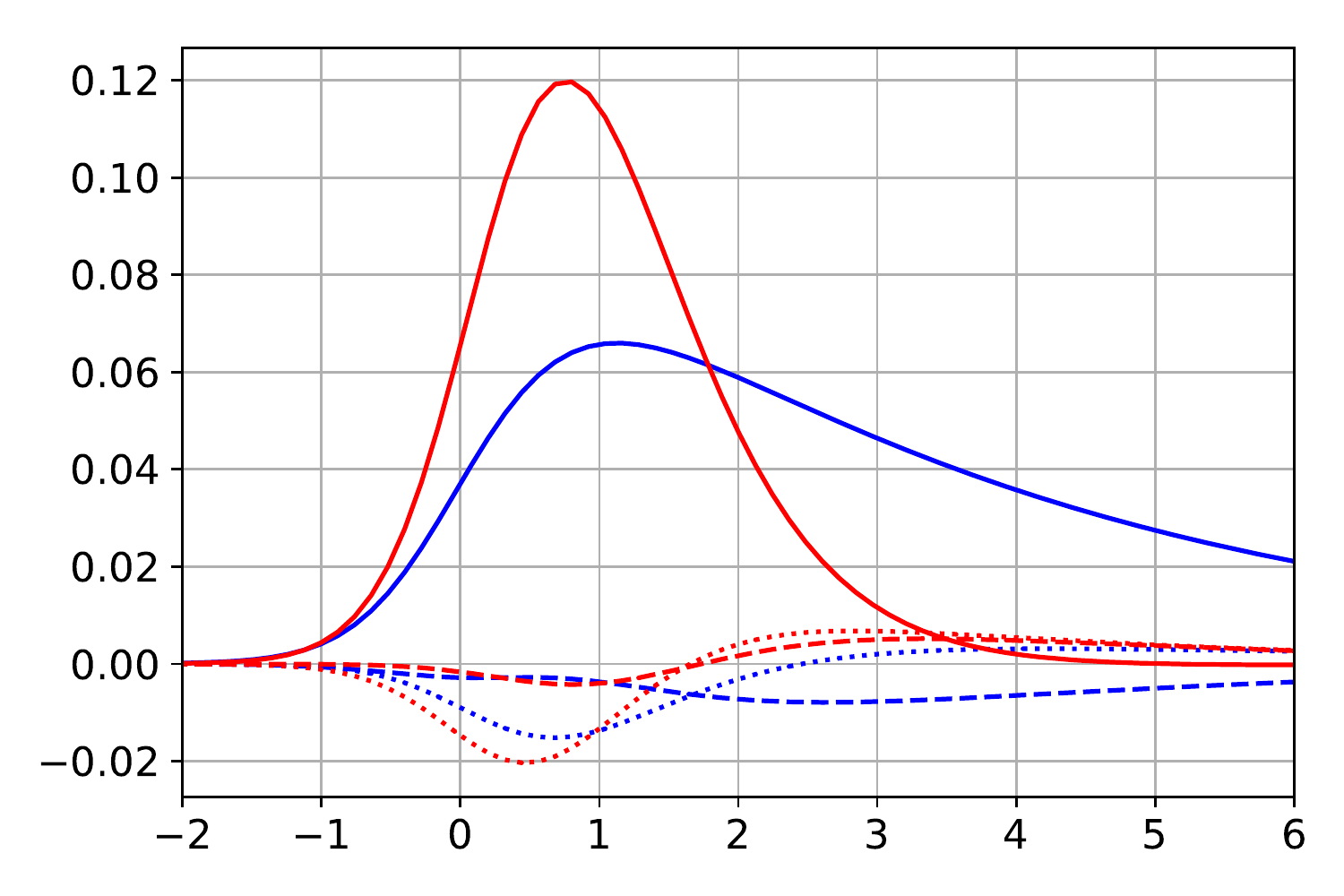}\includegraphics[scale=0.3]{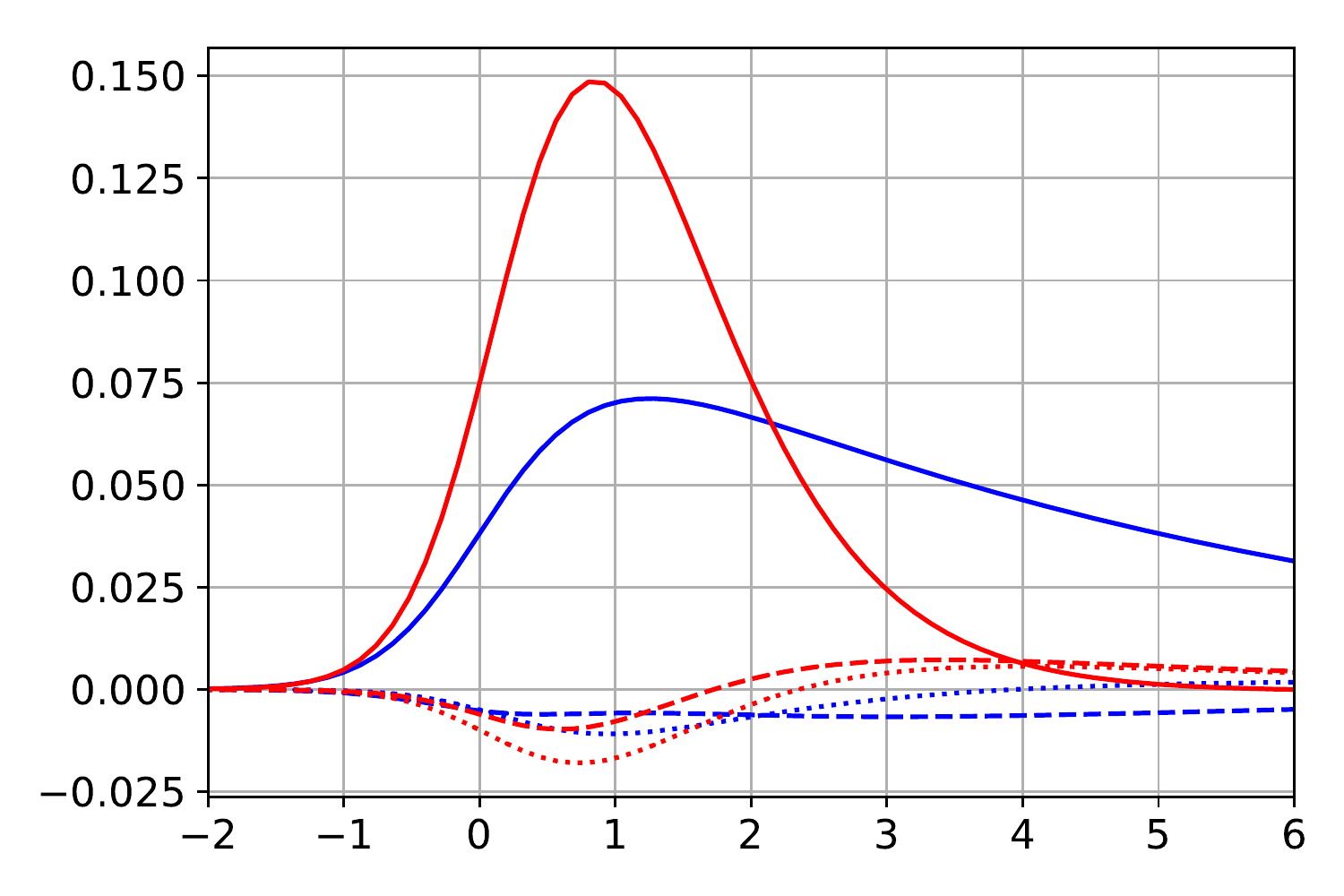}}
\centerline{$\!\!\!\!\!$\raisebox{0.5cm}{\rotatebox{90}{New Formalism}}\includegraphics[scale=0.3]{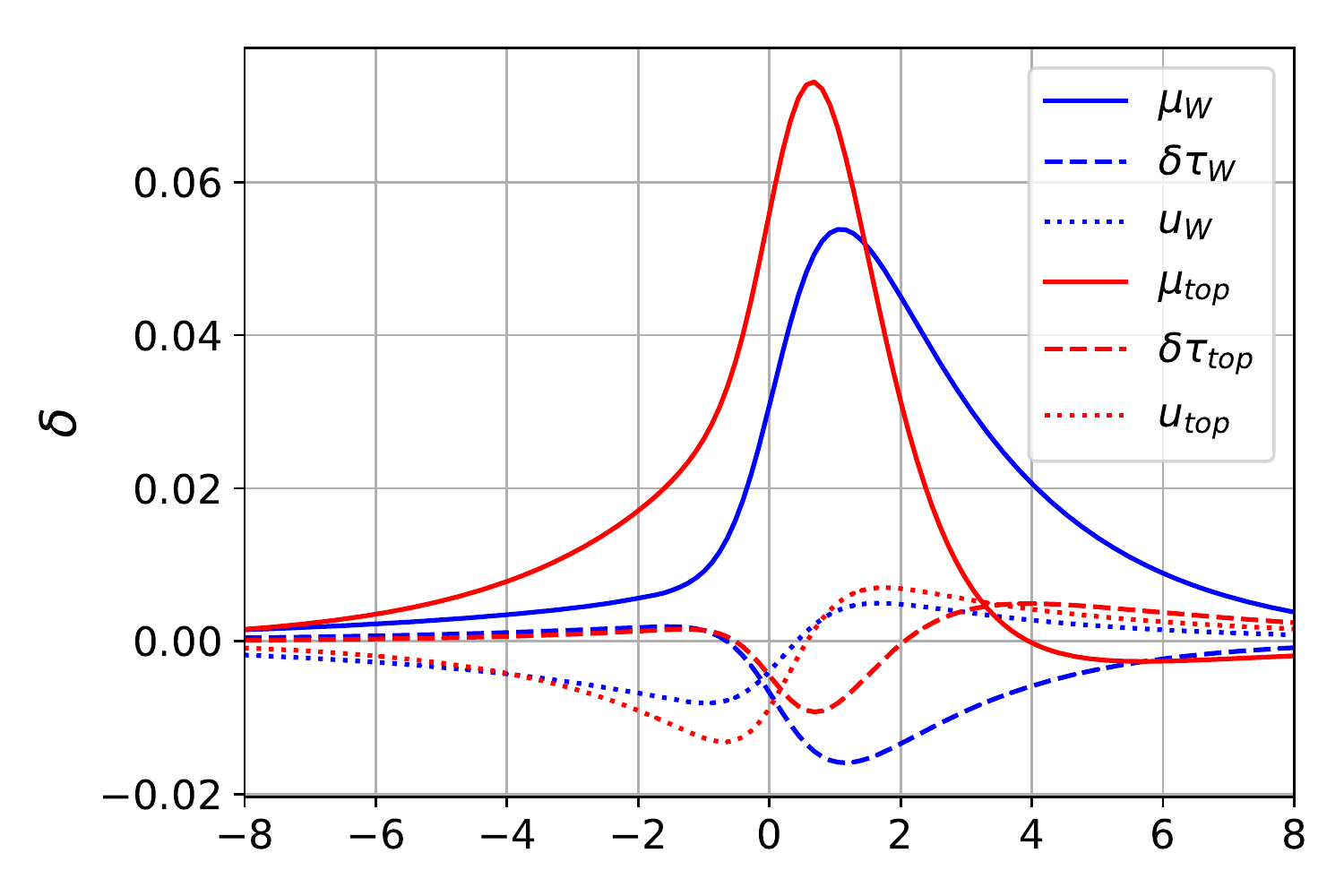}\includegraphics[scale=0.3]{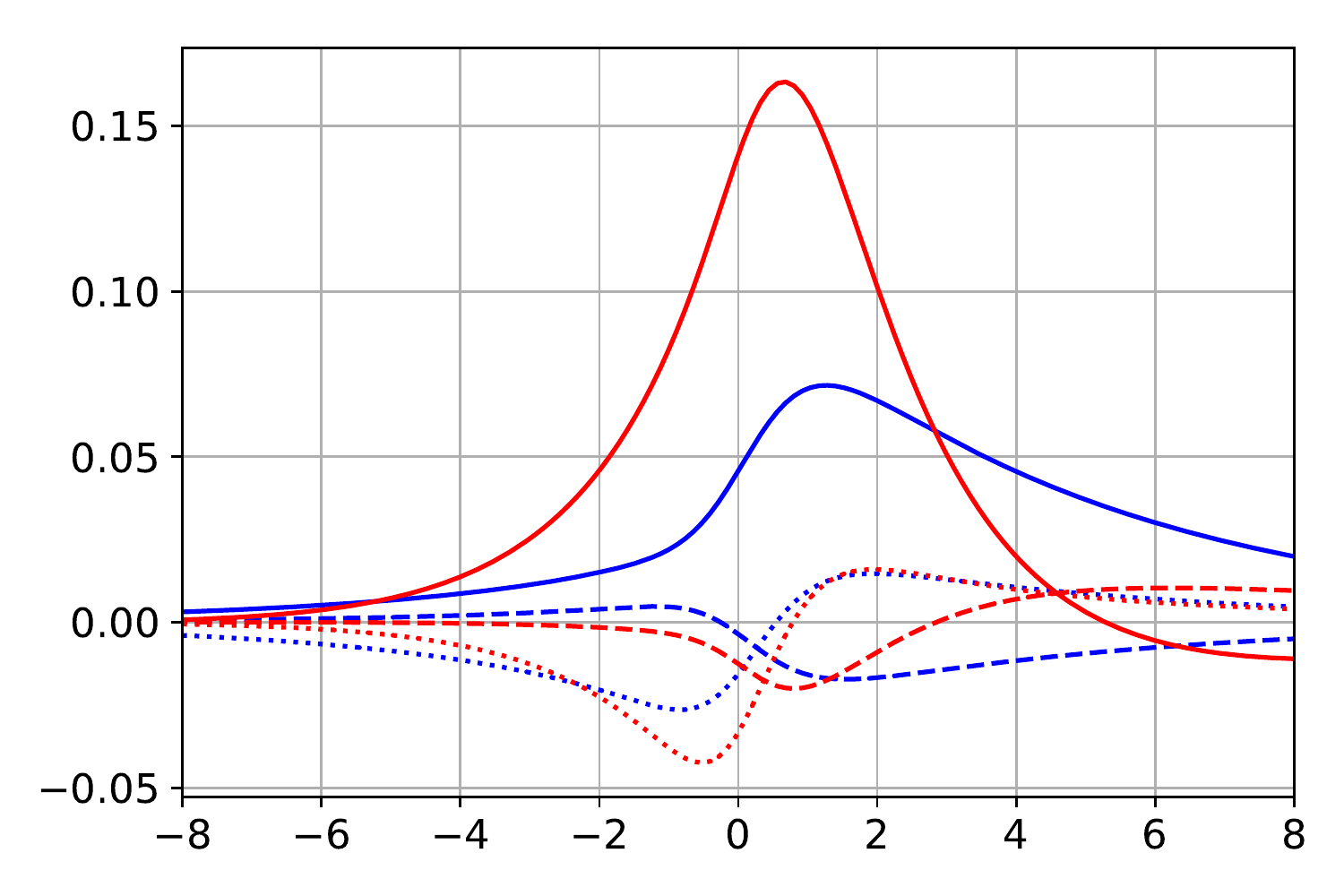}
\includegraphics[scale=0.3]{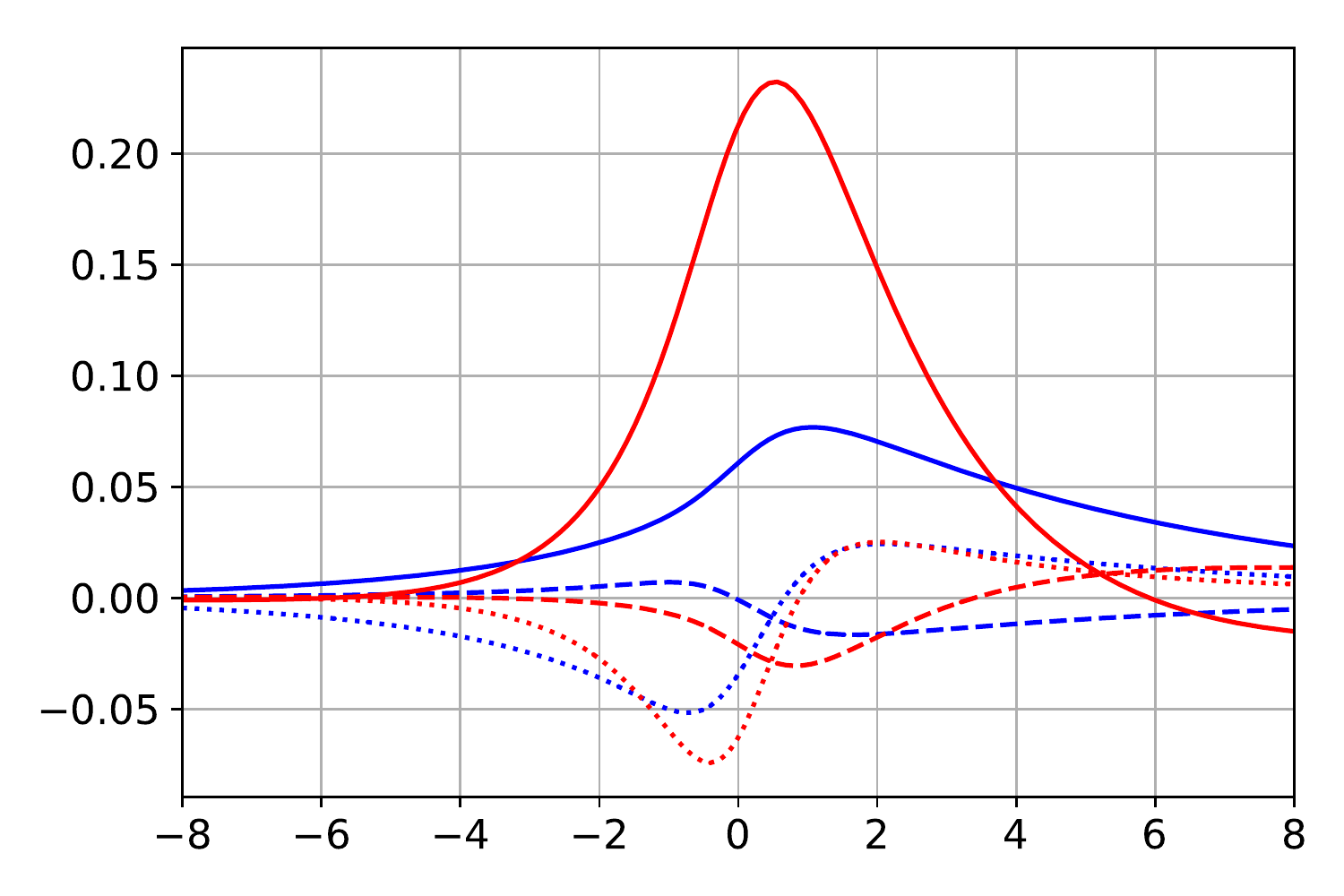}\includegraphics[scale=0.3]{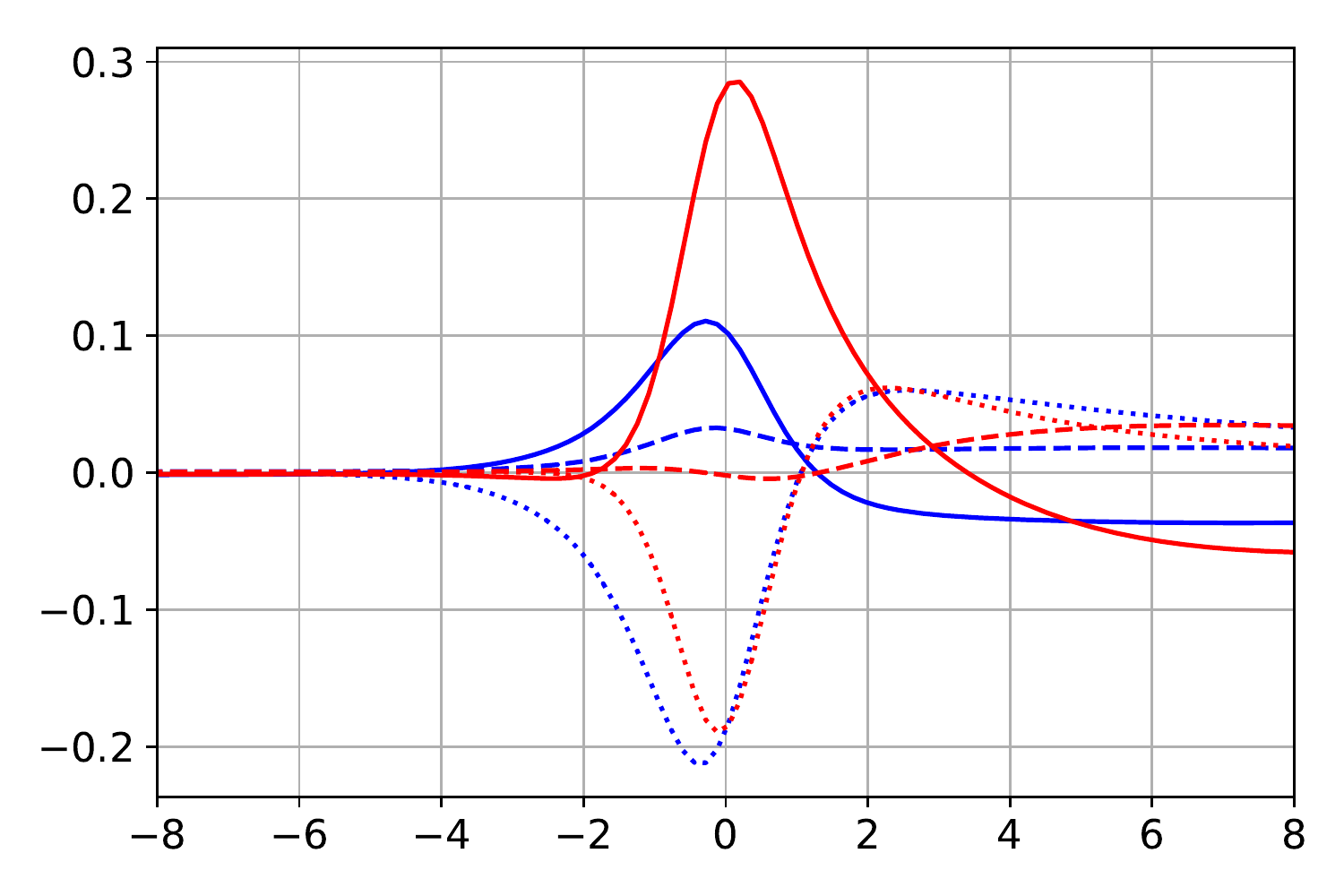}}
\centerline{$\!\!\!\!\!$\raisebox{1cm}{\rotatebox{90}{Friction}}\includegraphics[scale=0.3]{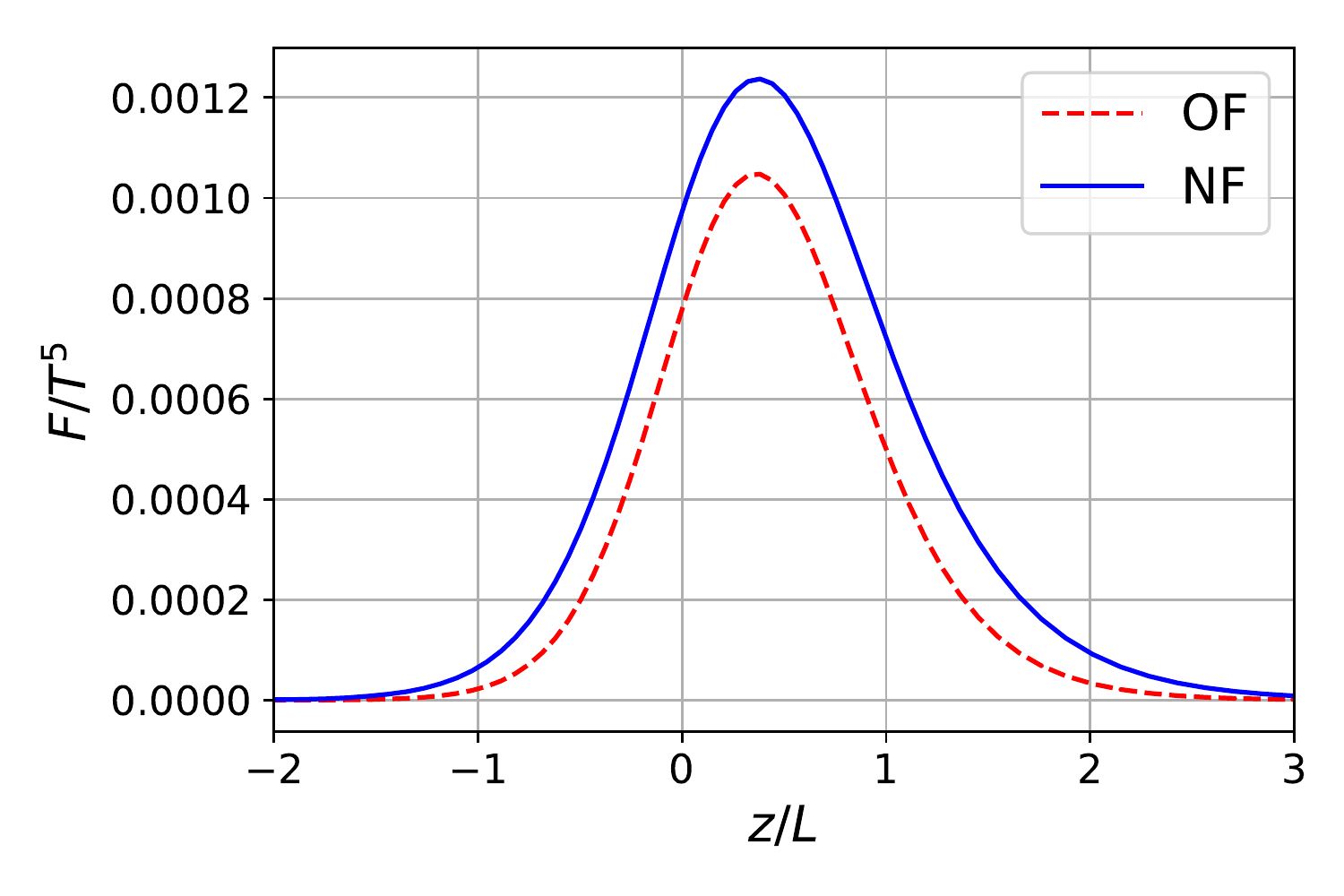}\includegraphics[scale=0.3]{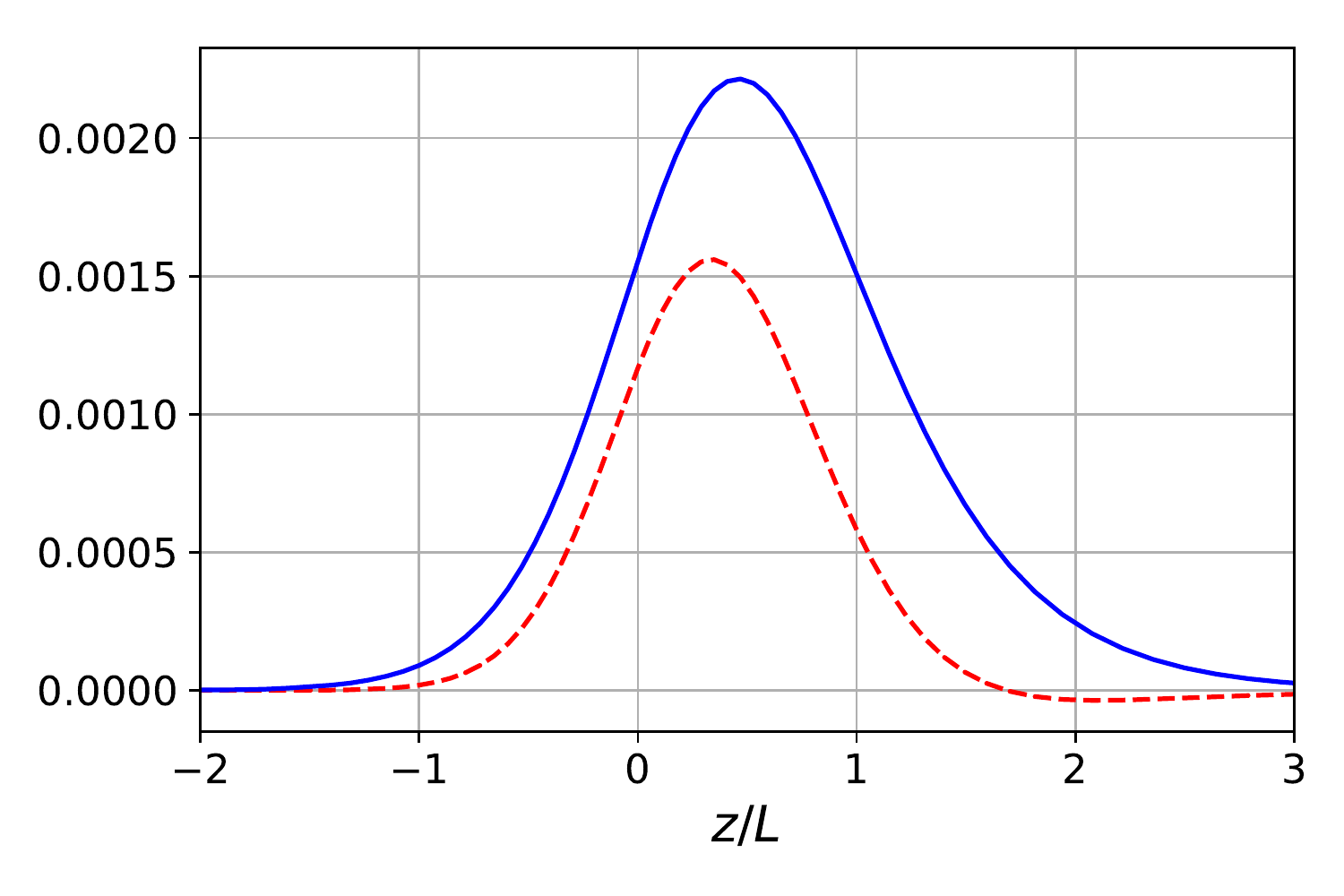}
\includegraphics[scale=0.3]{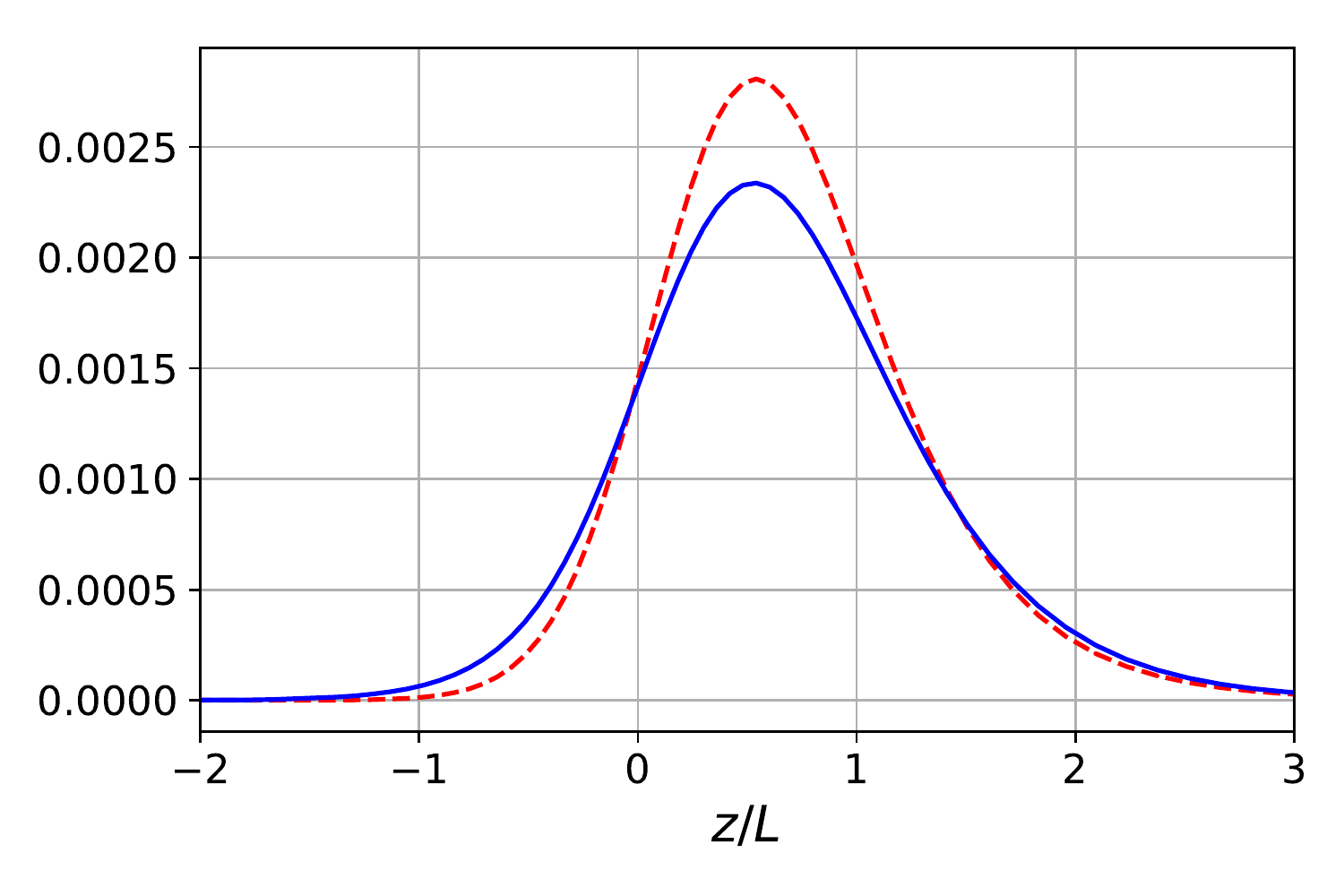}\includegraphics[scale=0.3]{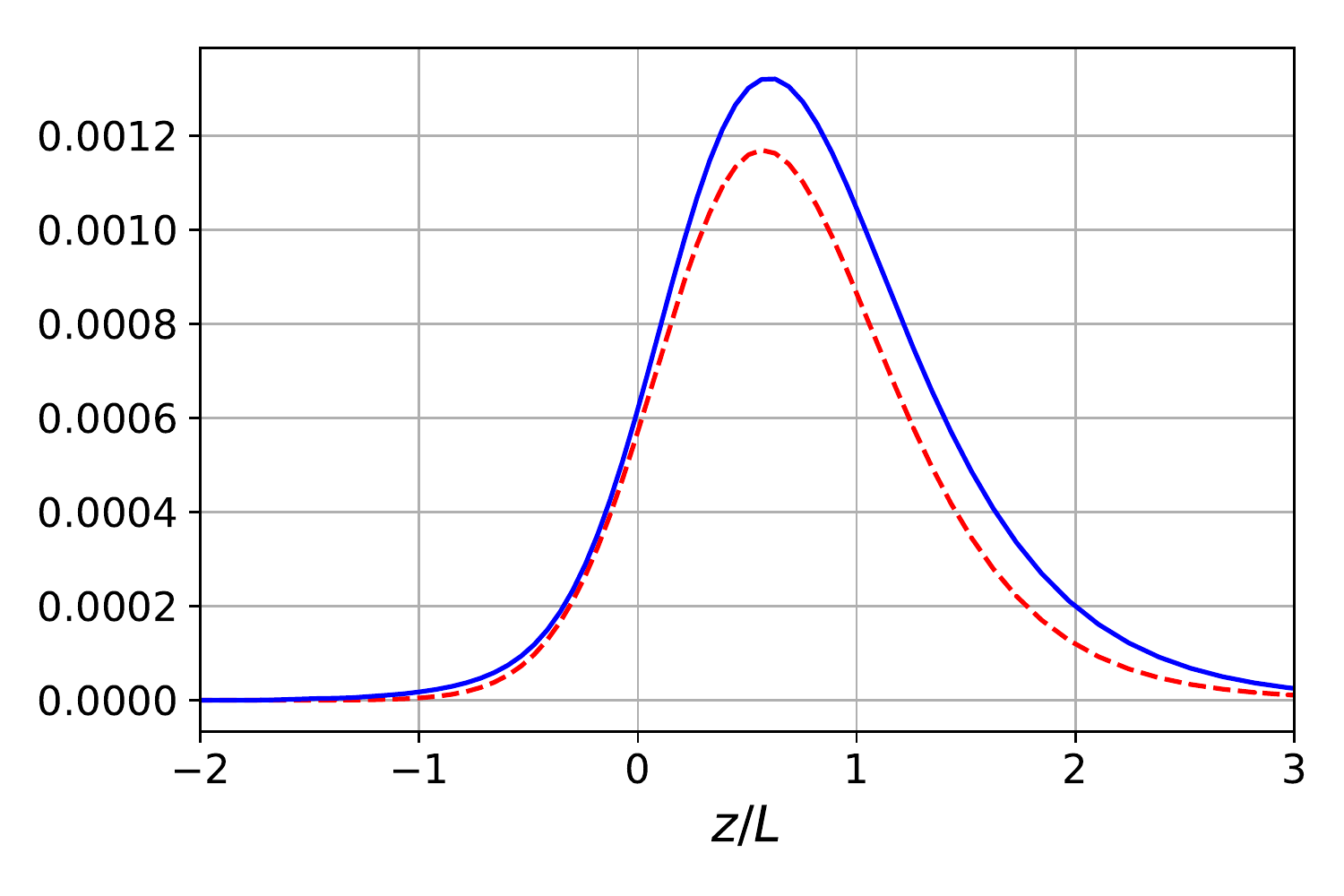}}
 \caption{First row: solutions for the perturbations of the $W$ and $t$ fluids within the old formalism, for $T=100\, \mathrm{GeV}$, $L=5\gamma /T$, $h_0=150\, \mathrm{GeV}$ and wall velocities $v = 0.2,\, 0.5,\, 0.7,\, 0.95$, as a function of $z/L$.
Second row: corresponding results for the improved fluid equations.  Third row: comparison of the friction term (\ref{friction}) obtained with both formalisms, with solid curves for NF and dashed for OF.  The symmetric phase in front of the bubble wall is to the left.}
 \label{fig:cont}
\end{figure*}


\section{Solutions for a Standard Model-like plasma}
\label{Solsect}
Next we apply
the improved fluid equations to a SM-like plasma in the context of a first order electroweak phase transition. The species that couple most strongly to the Higgs boson are the top quark $t$ and the electroweak gauge bosons. The $W$ and $Z$ bosons are approximated as having the 
same distribution functions, and we will refer to them collectively as W bosons. The remaining particles form a background fluid which is assumed to be in thermal equilibrium ($\mu_{\bg}=0$) at a $z$-dependent temperature $T+\delta\tau_{\bg}(z)$ \cite{Moore:1995si}. Even if they are not driven out of chemical equilibrium by the phase transition, these lighter fields still play an important role in the dynamics of the bubble wall. One might also expect the Higgs boson distribution to be perturbed, but its small number of degrees of freedom makes its contribution negligible compared to that of $t$ or $W$. It is therefore included in the background fluid (and similar reasoning could also
be applied to additional fields not present in the SM, {\it e.g.,} a singlet scalar). 

The complete set of matrix equations for the $t$, $W$
and background components is 
\bea
    A_t (q_t'+q_{\bg}') + \Gamma_t q_t &=& S_t \nn \\
    \label{syseq}
    A_\W (q_\W'+q_{\bg}') + \Gamma_\W q_\W &=& S_\W  \\
    A_{\bg} q_{\bg}' + \Gamma_{\bg,t} q_{t} + \Gamma_{\bg,\W} q_{\W} &=& 0 \nn 
\eea
where $A_t$, $A_\W$, $S_t$ and $S_\W$ are given in eqs.\ (\ref{Amatrix}) and (\ref{source}), using the appropriate equilibrium distribution functions. The $A$ matrix for the background fluid is 
\be
A_{\bg} = N_f A_t|_{m=0} + N_b A_\W|_{m=0},
\label{Abgeq}
\ee
with $N_f$ and $N_b$ respectively the fermionic and bosonic number of degrees of freedom included in the background fluid ($N_f=78$ and $N_b=19$ in the SM). We evaluate $A_t$ and $A_\W$ at $m=0$ because all the particles in the background fluid are approximately massless. Energy and momentum conservation fixes $\Gamma_{\bg,t} = -12\Gamma_t$ and $\Gamma_{\bg,\W} = -9\Gamma_\W$ \cite{Konstandin:2014zta}, and $\Gamma_t$ and $\Gamma_\W$ are evaluated in Appendix B.

\begin{figure*}[ht]
\centerline{\includegraphics[scale=0.28]{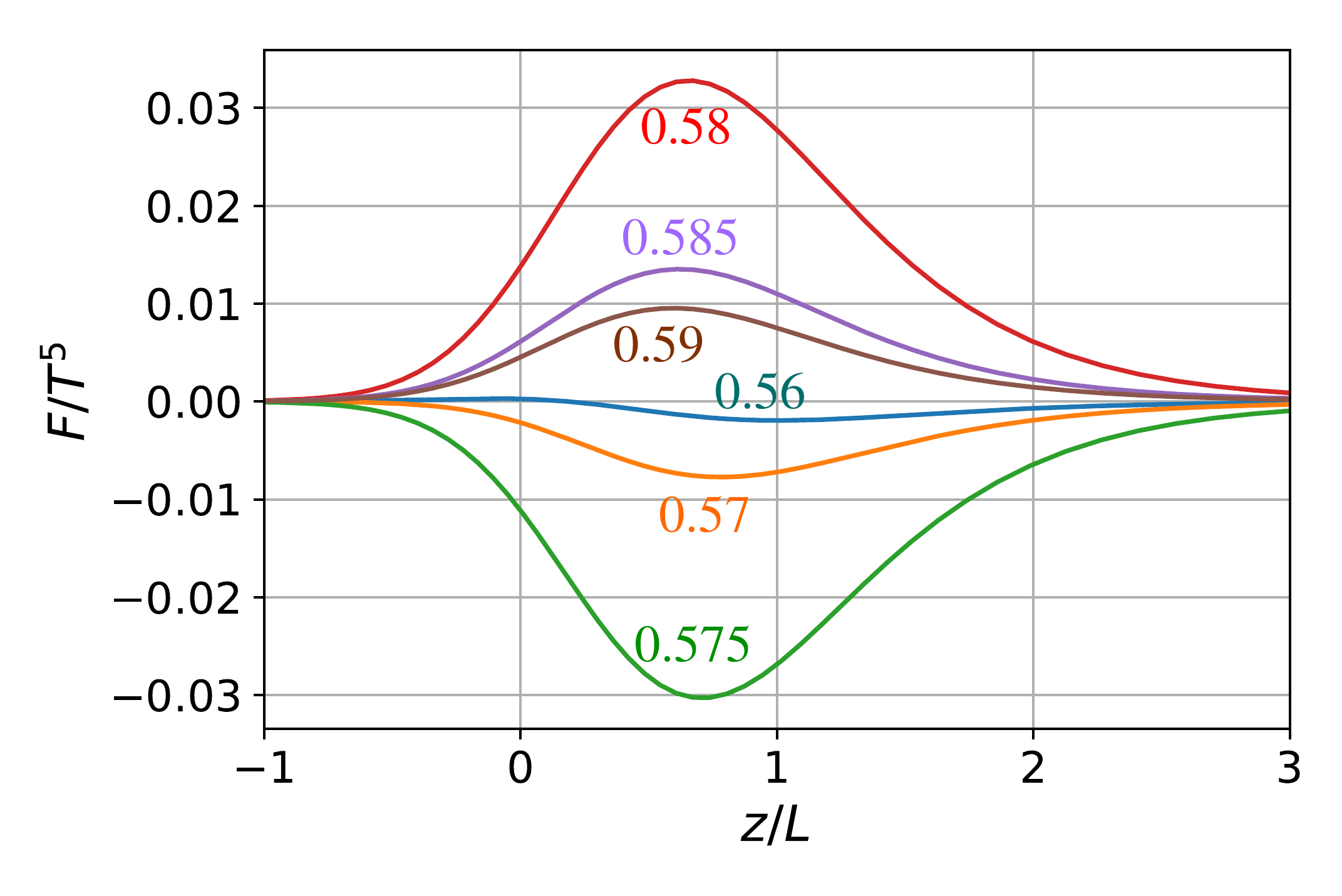}
\includegraphics[scale=0.4]{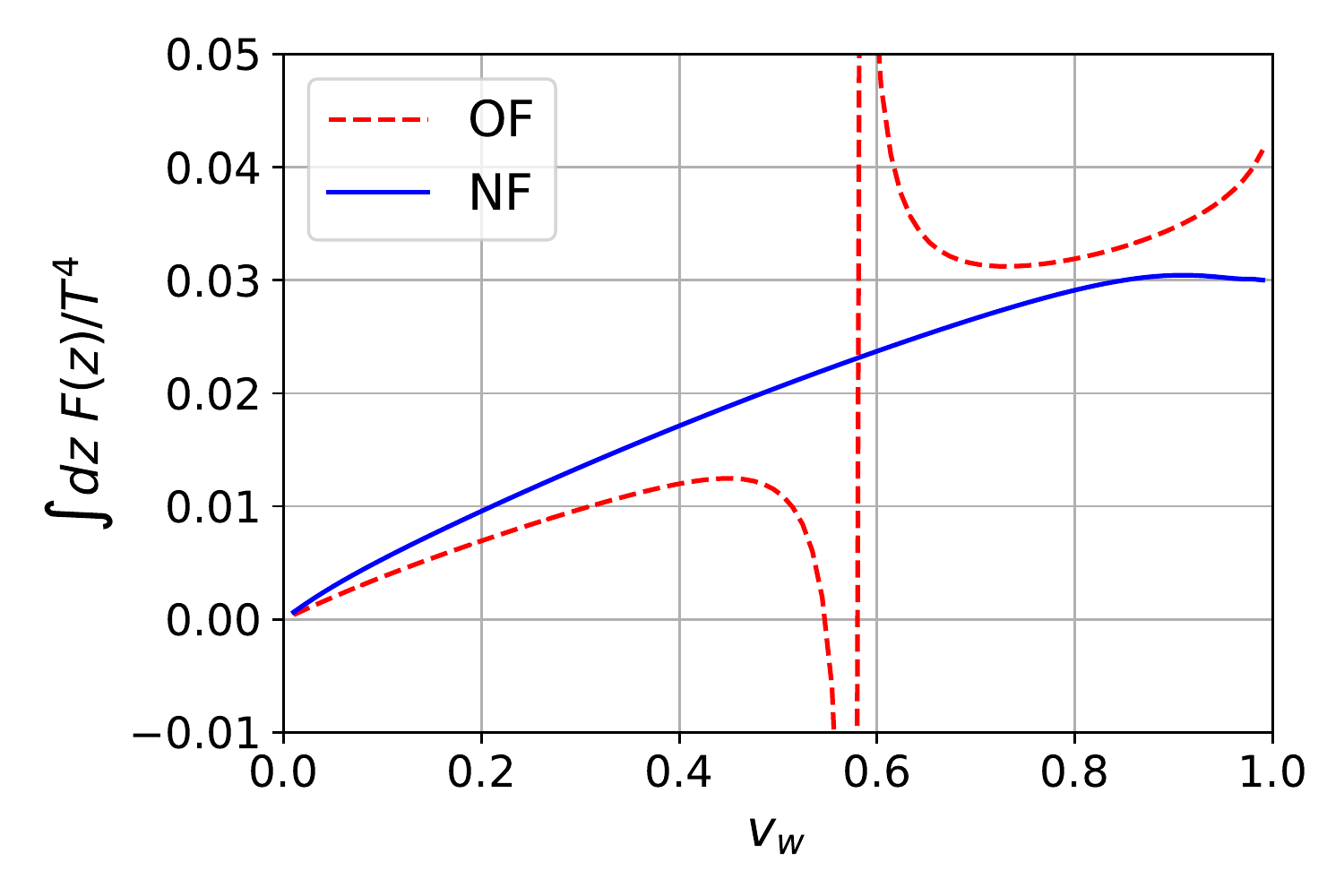}\includegraphics[scale=0.4]{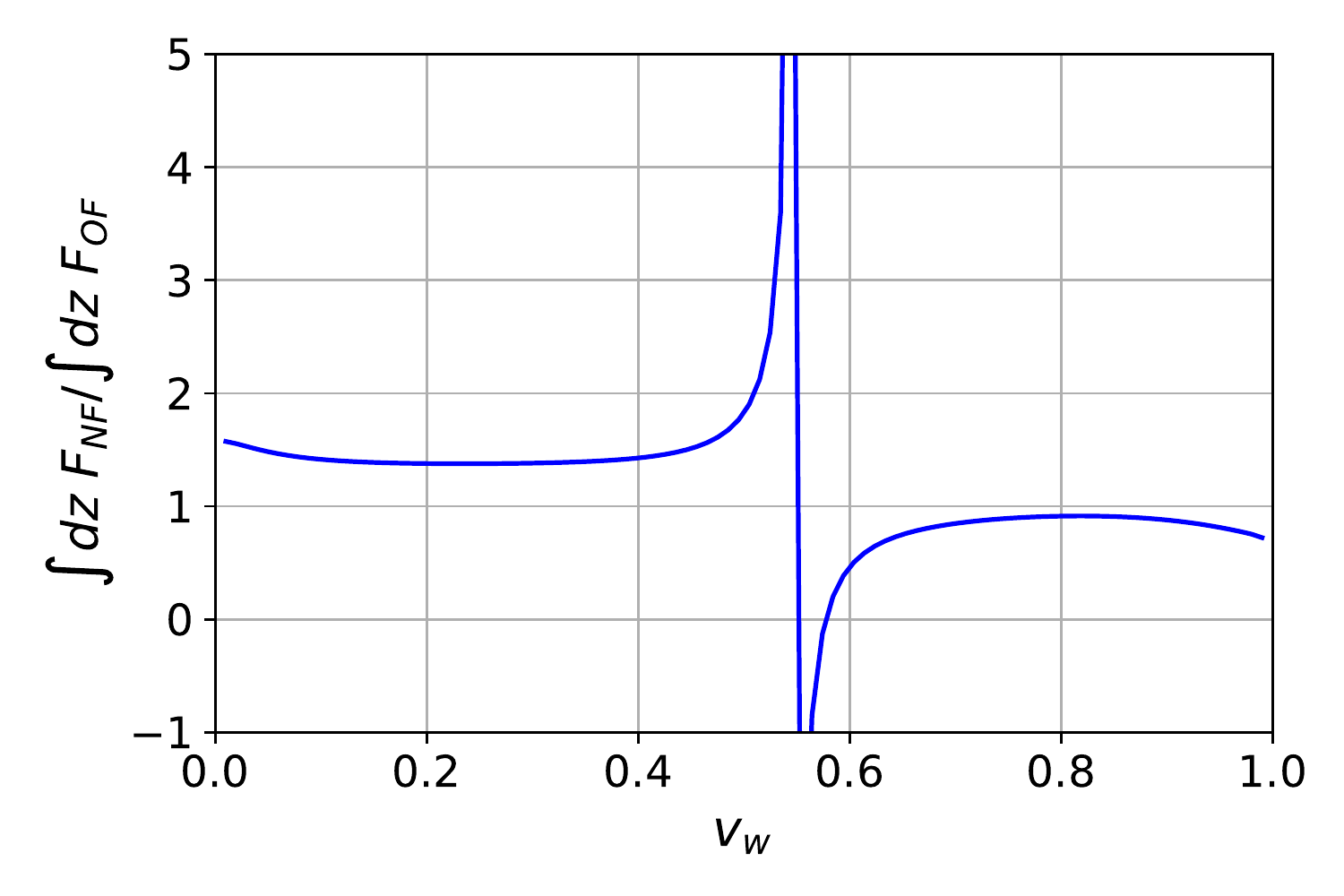}}
\vskip-0.3cm
\centerline{ (a) \qquad\qquad\qquad\qquad\qquad\qquad\qquad\qquad\qquad\qquad (b) \qquad\qquad
\qquad\qquad\qquad\qquad\qquad\qquad (c) \qquad\qquad\qquad}
 \caption{(a): Evolution of the friction with $v$ in the old formalism (OF), showing discontinuous behavior across the sound barrier. Each $F(z)$ curve is labeled by its value of $v$.  (b): The spatial integral of the friction in the OF (blue) and NF (orange) as a function of $v$, further illustrating the discontinuous behavior of the OF around the sound speed, and the smooth behavior of the NF. (c): The ratio of the two curves in (b). All the curves were obtained with $T=100\, \mathrm{GeV}$, $L=5/T$ and $h_0=150\, \mathrm{GeV}$.}
 \label{fig:cs}
\end{figure*}

To solve the system 
(\ref{syseq}), one can eliminate $q_{\bg}'$ using the third equation; however the fact that $\mu_{\bg}=0$ makes one of the three ``bg'' equations redundant. We have chosen to keep the first and third ``bg'' component equations (corresponding to the weighting factors $1$ and $p_z/E$), since this leaves $A_{\bg}$ nonsingular for $v\in (0,1)$. The result is
\be
    q_{\bg}' = -\tilde{A}_{\bg}^{-1} (\Gamma_{\bg,t}q_t + \Gamma_{\bg,\W}q_\W)
\ee
where $\tilde{A}_{\bg}^{-1}$ is the inverse of the $2\times 2$ $A_{\bg}$ matrix, projected onto the 1,3 columns and 2,3 rows of a $3\times 3$ matrix. It can be written in terms of the 3 matrices
\be
    P_1 = \left( \ba{ccc}
    0 & 0 & 0 \\
    0 & 1 & 0 \\
    0 & 0 & 1 \ea \right) ,\ 
    P_2 = \left( \ba{ccc}
    1 & 0 & 0 \\
    0 & 0 & 0 \\
    0 & 0 & 1 \ea \right) ,\ 
    P_3 = \left( \ba{ccc}
    0 & 0 & 0 \\
    1 & 0 & 0 \\
    0 & 0 & 0 \ea \right) \nn
\ee
and the $3\times 3$ $A_{\bg}$ matrix defined in (\ref{Abgeq}): 
\be
    \tilde{A}_{\bg}^{-1} = (P_2 A_{\bg} P_1 + P_3)^{-1} - P_3^\intercal
\ee

The six remaining equations take the form
\be
\label{beSM}
    A\,q' + \Gamma\, q = S
\ee
with
\bea
    A = \left( \ba{cc} A_\W & 0 \\ 0 & A_t \ea \right),\ S = \left( \ba{c} S_\W \\ S_t \ea \right),\ q = \left( \ba{c} q_\W \\ q_t \ea \right), \\
    \Gamma = \left( \ba{cc} 
    \Gamma_W - A_\W \tilde{A}_{\bg}^{-1} \Gamma_{\bg,\W} & -A_\W \tilde{A}_{\bg}^{-1} \Gamma_{\bg,t} \\ 
    -A_t \tilde{A}_{\bg}^{-1} \Gamma_{\bg,\W} & \Gamma_t - A_t \tilde{A}_{\bg}^{-1} \Gamma_{\bg,t} \ea \right)
\eea

To compare the new and old formalisms (denoted by NF and OF in the following) for a generic first order phase transition, we
model the bubble wall using a tanh ansatz for
the background Higgs field,
\be
      h(z) = \frac{h_0}{2}\big[1+\tanh(z/L)\big]
      \label{tanh_profile}
\ee
where $h_0$ is the VEV of the Higgs in the broken phase and $L$ is the wall thickness. As an example we solve  eqs. (\ref{beSM}) within the OF and NF for $T=100\, \mathrm{GeV}$, $h_0=150\, \mathrm{GeV}$, $L/\gamma=5/T$ \footnote{$L/\gamma$ is the wall thickness as measured in the plasma frame.  Fixing $LT/\gamma$ rather than $LT$ makes it easier to see that the  diffusion tails in front of the wall disappear as $v\to 1$.} and several wall velocities, using the collision rates given in \cite{Moore:1995si} for the OF and the ones evaluated in Appendix B for the NF. We include a factor $\gamma$ in $L$ in order for the wall to have a constant thickness in the plasma frame. The solutions are shown in Figure \ref{fig:cont}, for a series of 
increasing wall velocities.

One can notice that within the NF, the perturbations in front of the wall ($z<0$) vanish only in the limit $v \to 1$, as required by causality. This is not the case in the OF, whose solutions always vanish in front of the wall for $v>1/\sqrt{3}$. As argued in ref. \cite{Cline:2020jre}, this behavior is unphysical, since there is no reason for particles not to be able to diffuse in front as long as their $v_z$  velocity
component is higher than $v$.

As a consistency check, we observe that the linearization of the Boltzmann equation in $\delta X$ and $u$ is justified, since all the perturbations are generally well below unity in magnitude. We have tested that this condition holds for most wall parameters; the linearization starts to break down only in the extreme cases of very fast ($v\gtrsim 0.95$) and thin walls ($L\lesssim 1/T$). 

\section{Consequences for wall friction}
\label{fric-sect}

An important application is the calculation of the friction term $F$ in the Higgs equation of motion multiplied by $h' = dh/dz$ \cite{Konstandin:2014zta},
\be
E_h \equiv h'' h' - \left.\partial V_{\rm eff}\over\partial z\right|_T - F = 0,
\label{higgs_eom}
\ee
where $V_{\rm eff}|_T$ is the finite-temperature potential evaluated at the unperturbed background temperature, and 
\bea
\label{friction}
    F(z) &=& \sum_i \frac{dm_i^2}{dz} N_i \int \frac{d^3p}{(2\pi)^3 2E}(\delta\! f_{u,i}-f_{v,i}' \delta \!\bar{X}_i) \nn \\
    &=& \sum_i \frac{dm_i^2}{dz}\frac{N_i T^2}{2} \Big[ C_0^{1,0} \mu_i + C_0^{0,0}(\delta\tau_i+\delta\tau_{bg})  \nn\\
    && +  D_v^{0,-1}(u_i+u_{bg})\Big] \,.
\eea
Here the sum is over the species $t$ and $W$, and $N_i$ is the corresponding number of degrees of freedom.  An exact solution to eq.\  (\ref{higgs_eom}) exists only for a specific wall velocity and shape, and so the accurate estimation of $F$ is important for determining the wall properties.  An ansatz such as (\ref{tanh_profile}) can give a rough approximate solution, where $v$ and $L$ are determined by demanding that two moments of eq.\ 
 (\ref{higgs_eom}) vanish \cite{Moore:1995si,Konstandin:2014zta,Kozaczuk:2015owa}, for example
 \bea
 M_1 &\equiv& \int dz\, E_h=0,\nn\\
 M_2 &\equiv& \int dz\, E_h (2h - h_0)=0
 \label{EOMmom}
 \eea

We plot $F(z)$ constructed from the OF and NF solutions in the bottom row of Figure \ref{fig:cont}.
At small $v$, the friction predicted by NF is $\sim 20\%$ larger,
leading us to expect the NF to predict a smaller wall velocity than the OF for subsonic walls. This difference is mainly due to our improved calculation of the collision integrals and the fact that we keep the full mass dependence of the $C_v^{m,n}$ and $D_v^{m,n}$ functions. In this very coarse grid on velocity space, $v = 0.2,$ $0.5$, $0.7$, $0.95$, the friction appears to be qualitatively similar in shape at each velocity, involving primarily a modest rescaling factor to relate the results of the two approaches.

Despite the appearance in Fig.\ \ref{fig:cont}
of no dramatic difference between the two formalisms, more careful investigation 
in the vicinity of the sound speed reveals
the crucial pathology of the OF.  In Fig.\ 
\ref{fig:cs}(a) we plot $F(z)$ for a series
of wall speeds from $0.56$ to $0.59$ within
the OF, revealing that it briefly becomes
{\it negative} before suddenly becoming positive again.  This is even more clear 
in terms of the integral of the friction $\int dz\, F(z)$, which we plot as a function of $v$
for the  OF and NF in Fig.\ \ref{fig:cs}(b).
The integral undergoes a discontinuity near
$v=c_s$ in the OF, while remaining smooth
and continuous in the NF.  The ratio of
the integrals between the NF and OF is plotted
in fig.\ (\ref{fig:cs})(c), underscoring the relatively good two agreement of the two, except close to $c_s$.

In addition to giving incorrect results close
to $v=c_s$, 
this discontinuous behavior of the OF makes it
difficult to automate searches for wall properties, since the jump in $\int F\,dz$ leads to a similar discontinuity in the moment $M_1$ whose zero is being searched for.  
As expected, the second moment $M_2$ is also discontinuous in the OF.  This was the practical difficulty that prompted our investigation.
In contrast the NF gives smooth results,
which we have argued is the expected behavior on physical grounds, since diffusion should not be greatly sensitive to  whether the
speed is slightly above or below $c_s$.

\section{Conclusion}
\label{conc}

In this work we have pointed out a shortcoming at high wall speeds ($v\gtrsim c_s$) with the
 fluid equations that have been used, since their introduction in ref.\ \cite{Moore:1995si}, to calculate the friction $F$ on electroweak bubble walls. We have also proposed a modification to these
 equations that solves the problem.  It is reassuring that the two approaches give results that are not too different from each other at 
 low wall speeds---and there the difference arises mainly because we have improved estimates of the collision rates, rather than the changes in formalism that become important at high $v$.  Near the sound barrier and above, the differences are more significant, with our new results evolving continuously as a function of $v$, whereas the old ones exhibit a discontinuity in $F$ at $v = c_s$.
 The new system predicts lower friction at high 
 $v>c_s$ compared to the old one, which is likely to 
 lead to faster walls.  At low $v$ the opposite is true.  Application of these methods to a realistic
 model is underway \cite{Cline:2021iff}.

The new elements in our treatment are a different choice of weighting factors for taking moments of the Boltzmann equation, and a different treatment of the velocity perturbation. The latter has long been recognized and recently highlighted in the high-$v$ context in ref.\ \cite{{Cline:2020jre}}.  While there are strong
theoretical motivations for the velocity perturbation,
the choice of weighting factors is more arbitrary,
and cannot be justified {\it a priori}. 

Instead we have made a phenomenological
determination, by finding a set of 
moments that give the expected behavior
for the fluid perturbations as a function of $v$.  One could characterize it as an educated guess, that should be validated by finding a more exact solution of the full Boltzmann equations.  There are several ways one could imagine doing this.  Instead of three moments and three perturbations, one could increase this number to $N$ and look for convergence of a physical quantity like the friction with increasing $N$.  Alternatively, one could approximate the distribution function $f$ by taking $N$ bins in momentum space and seeking convergence
with growing $N$.  This is an investigation we hope to undertake in future work.

{\bf Acknowledgment.}  We thank A.\ Friedlander,
K.\ Kainulainen and D.\ Tucker-Smith for useful discussions.
This work was supported by NSERC (Natural Sciences
and Engineering Research Council, Canada)
and FRQNT (Fonds de recherche Nature et technologies, Qu\'ebec).
\begin{appendix}
\section{$v$-dependence of the $C_v^{m,n}$ and $D_v^{m,n}$ functions}\label{appA}
The coefficients appearing in the $A$ matrix generally depend on the local particle masses $m(z)/T$ and the wall velocity $v$. They can be evaluated numerically directly from their definition (\ref{Cmn}), but it is also possible to analytically 
calculate their $v$-dependence, by making the substitution $E\to \gamma(E+v p_z)$ and $p_z\to \gamma(p_z+v E)$ to boost the integration variables to the plasma frame. This  transforms $f_v$ to $f_0$, the equilibrium distribution function evaluated at $v=0$, and leaves the combination $d^3p/E$ invariant. 

In this way, the $C_v^{m,n}$ and $D_v^{m,n}$ functions can be expressed as a sum (finite or infinite) of $C_0^{m,n}$ and $D_0^{m,n}$, the corresponding functions evaluated at $v=0$. One can show that (henceforth omitting the subscript $0$)
\bea
    C_v^{-1,1} &=& \gamma^3 v\,[C^{-2,0}+(2+v^2)C^{0,2}] \nn \\
    C_v^{0,0} &=& \gamma\, C^{0,0} \nn \\
    C_v^{0,1} &=& \gamma^2 v\,(C^{-1,0}+C^{1,2}) \nn \\
    C_v^{0,2} &=& \gamma^3\, [v^2 C^{-2,0}+(1+2v^2)C^{0,2}] \nn \\
    C_v^{1,0} &=& C^{1,0} \nn \\
    C_v^{1,1} &=& \gamma v\, C^{0,0} \nn \\
    C_v^{1,2} &=& \gamma^2 (C^{1,2}+v^2 C^{-1,0}) \nn \\
    C_v^{2,1} &=& v\, C^{1,0}-\frac{1}{\gamma^2}\sum_{n=1}^\infty v^{2n-1} C^{2n+1,2n} \nn \\
    C_v^{2,2} &=& \gamma v^2 C^{0,0}+\frac{1}{\gamma^3}\sum_{n=1}^\infty v^{2n-2} C^{2n,2n} \nn
\eea
\bea
    C_v^{2,3} &=& \gamma^2 v^3 C^{-1,0}+\gamma^2 v(v^4-3v^2+3)\,C^{1,2} \nn\\ &\ &-\frac{1}{\gamma^4}\sum_{n=2}^\infty v^{2n-3} C^{2n-1,2n} \nn \\
    D_v^{-1,0} &=& \gamma^2 (D^{-1,0}+v^2 D^{1,2}) \nn \\
    D_v^{0,0} &=& \gamma D^{0,0} \nn \\
    D_v^{1,1} &=& \gamma v D^{0,0} \nn 
\eea
With these, it is sufficient to compute the required $C^{m,n}$ and $D^{m,n}$ at only a few values of $m/T$ and use interpolation to quickly compute them for any $m/T$. The infinite series are all well-behaved: they are exact at $v=0$ and $v=1$ using only the first term of the series, and  an accuracy of less than 1\% for
all $v\in [0,1]$ is achieved using a small number of terms.\\

\section{Evaluation of the collision rates}
\label{appB}
We discuss here the calculation of the collision integrals by a corrected and improved version of the method used in ref. \cite{Moore:1995si}.
The collision term for a given particle species is
\bea
   \mathcal{C}[f_v(p)] &=& \sum_i \frac{1}{2N_pE_p} \int\frac{d^3k\, d^3p'\, d^3k'}{(2\pi)^5 2E_k 2E_{p'} 2E_{k'}}\,\vert\mathcal{M}_i\vert^2\nn\\
   &\times& \delta^4(p+k-p'-k')\,\mathcal{P}[f_v(p)]\,;\\
  \mathcal{P}[f(p)] &=& f(p)f(k)\big(1\pm f(p')\big)\big(1\pm f(k')\big)\nn\\ &-&f(p')f(k')\big(1\pm f(p)\big)\big(1\pm f(k)\big)\,,
\eea
where the sum is over all the relevant processes listed in Table
\ref{amplitudes}, $p$ is the momentum of the incoming particle
whose distribution is being computed, $N_p$
is its number of degrees of freedom,
$k$ is the momentum of the other incoming particle, and $p'$, $k'$ are the momenta of the outgoing particles. $|\mathcal{M}_i|^2$ is the squared scattering amplitude, summed over the helicities and colors of all the external particles. The distribution functions appearing in $\mathcal{P}$ are Fermi-Dirac
or Bose-Einstein depending the respective external particles,
and  the $\pm$ is $+$ for bosons and $-$ for fermions.

$\mathcal{P}$ can be simplified by expanding it to linear order in the perturbations. Using the definition (\ref{dfeq}) of the distribution function with $\delta X(p) = \mu+\beta\gamma\delta\tau(E_p-v p_z)-\delta f/f_v'$, one can show that $\mathcal{P}$ becomes
\be
\label{linearPop}
    \mathcal{P}[f] = f(p)f(k)(1\pm f(p'))(1\pm f(k'))\sum (\pm\delta X)
\ee
where the sum is over the external particles not in equilibrium and the $\pm$ in front of $\delta X$ is $+$ for incoming particles and $-$ for outgoing particles.

The quantities needed for the fluid equations are the moments of $\mathcal{C}[f]$. These have the general form
\bea
\label{coll}
 \sum_i \frac{1}{2N_pE_p}&&\!\!\!\!\!\!\!\! \int\frac{d^3k\, d^3p'\, d^3k'}{(2\pi)^5 2E_k 2E_{p'} 2E_{k'}}\,\vert\mathcal{M}_i\vert^2\nn\\
   &\times& \delta^4(p+k-p'-k')\,\mathcal{P}[f_v]\,\frac{p_z^n}{E_p^m}\\ 
=\sum_i \frac{1}{2N_pE_p}&&\!\!\!\!\!\!\!\! \int\frac{d^3k\, d^3p'\, d^3k'}{(2\pi)^5 2E_k 2E_{p'} 2E_{k'}}\,\vert\mathcal{M}_i\vert^2\nn\\
   \!\!\!\!\!\!\!\!\times\ \ &&\!\!\!\!\!\!\!\! \delta^4(p+k-p'-k')\,\mathcal{P}[f_v]\,\gamma^{n-m}\frac{(p_z+v E_p)^n}{(E_p+v p_z)^m}\nn
  \eea
where we boosted to the plasma frame to get the second line. Using the substitution (\ref{substitution}), the perturbations become in that frame
\be
    \delta X(p) = \mu + \beta E_p \delta\tau - \left(\frac{E_p+v p_z}{p_z+v E_p}\right) \left(\frac{f_0}{f_0'}\right) u
\ee
\begin{table}[t!]
\centering
\begin{tabular}{|c|c|}
    \hline
    Process & $\vert\mathcal{M}\vert^2$ \\
    \hline
    {\bf Top quark:} & \\
    $\bar{t}t \to gg$ & $-\frac{128}{3} g_s^4 \frac{st}{(t-m_q^2)^2}$\\
    $tg \to tg$ & $-\frac{128}{3} g_s^4 \frac{su}{(u-m_q^2)^2} + 96g_s^4 \frac{s^2+u^2}{(t-m_g^2)^2}$ \\
    $tq \to tq$ & $160g_s^4 \frac{s^2+u^2}{(t-m_g^2)^2}$ \\
    \hline
    {\bf W bosons:} & \\
    $Wq \to qg$ & $-72g_s^2 g_w^2 \frac{st}{(t-m_q^2)^2}$\\
    $Wg \to \bar{q}q$ & $-72g_s^2 g_w^2 \frac{st}{(t-m_q^2)^2}$\\
    $WW \to \bar{f}f$ & $-\frac{27}{2}g_w^4 st \left[\frac{3}{(t-m_q^2)^2} + \frac{1}{(t-m_l^2)^2}\right]$ \\
    $Wf \to Wf$ & $360g_w^4 \frac{u^2}{(t-m_W^2)^2} - \frac{27}{2}g_w^4 su \left[\frac{3}{(u-m_q^2)^2} + \frac{1}{(u-m_l^2)^2}\right]$ \\
    \hline
\end{tabular}
\caption{Relevant processes for the top quark and W bosons and their corresponding scattering amplitude in the leading log approximation.}
\label{amplitudes}
\end{table}

Following the treatment of ref.\ \cite{Moore:1995si}, the calculation of the collision rates has been done to leading log accuracy, where it is justified to neglect the masses of all the
external particles, which implies $E_p=p$. One can also neglect $s$-channel contributions and the interference between diagrams because they are not logarithmic. To account for thermal effects, we use propagators of the form $1/(t-m^2)$ or $1/(u-m^2)$, where $m$ is the exchanged particle's thermal mass. It is given by $m_g^2=2g_s^2 T^2$ for gluons, $m_q^2=g_s^2 T^2/6$ for quarks, $m_W^2=5g_w^2 T^2/3$ for W bosons and $m_l^2=3g_w^2 T^2/32$ for leptons
\cite{Weldon:1982bn}.

The top quark collisions are dominated by their strong interactions; we include only contributions to
$|{\cal M}|^2$ of order $g_s^4$ for $t$
interactions.  For the $W$ bosons, we include 
terms of order  $g_s^2 g_w^2$ and $g_w^4$. The relevant processes are shown with their corresponding $|{\cal M}|^2$  in Table \ref{amplitudes} \footnote{As pointed out in ref.\ \cite{Arnold:2000}, there were some errors in the expressions of the scattering amplitudes in \cite{Moore:1995si}. They failed to include a $1/2$ symmetry factor in the amplitude for $\bar{t}t \to gg$ and made some algebraic errors in $tq \to tq$ and $Wf \to Wf$.}.

To evaluate the integrals in (\ref{coll}), one can first use the delta function and the symmetry of the integrand to  analytically
perform five of the twelve integrals. This can be done efficiently using the parametrization detailed in refs.\ \cite{Arnold:2000,Arnold:2003,Moore:2001}. The remaining seven integrals can be evaluated analytically using several approximations, justified to leading log accuracy. However, we have found that it is more precise to numerically compute  these integrals, which can be done with a Monte Carlo algorithm. One can use a stratified sampling algorithm or VEGAS to reduce the variance, but this is generally not necessary since it only takes a few seconds to get an accuracy of $\sim 1\%$
in most cases.

With the linearization of $\mathcal{P}[f]$ made in (\ref{linearPop}), the moments of the collision term can be written as linear combinations of the three perturbations: $T\left(\Gamma_\mu^{(i)} \mu + \Gamma_\tau^{(i)} \delta\tau + \Gamma_u^{(i)} u \right)$. Then the $\Gamma$ matrix appearing in eq.\ (\ref{matbe}) takes the form
\be
    \Gamma = T\left( \ba{ccc}
    \Gamma_\mu^{(1)} & \Gamma_\tau^{(1)} & \Gamma_u^{(1)} \\
    \Gamma_\mu^{(2)} & \Gamma_\tau^{(2)} & \Gamma_u^{(2)} \\
    \Gamma_\mu^{(3)} & \Gamma_\tau^{(3)} & \Gamma_u^{(3)}
    \ea \right)
\ee
where the $\Gamma_i^{(j)}$ coefficients are dimensionless.
The $v$-dependence of the upper-left $2\times 2$ block can be expressed analytically, giving
\bea
    \Gamma_{\mu,t}^{(1)} &=& 0.00196, \ \ \ \Gamma_{\mu,W}^{(1)} = 0.00239 \nn \\
    \Gamma_{\tau,t}^{(1)} &=& 0.00445, \ \ \  \Gamma_{\tau,W}^{(1)} = 0.00512 \nn \\
    \Gamma_{\mu,t}^{(2)} &=& 0.00445\,\gamma, \ \Gamma_{\mu,W}^{(2)} = 0.00512\,\gamma \nn \\
    \Gamma_{\tau,t}^{(2)} &=& 0.0177\,\gamma, \ \ \  \Gamma_{\tau,W}^{(2)} = 0.0174\,\gamma \nn
\eea
The remaining components have been fitted to quartic polynomials:
\bea
    \Gamma_{u,t}^{(1)} &=& (5.36v - 4.49v^2 + 7.44v^3 - 5.90v^4)\times 10^{-3} \nn \\
    \Gamma_{u,W}^{(1)} &=& (4.10v - 3.28v^2 + 5.51v^3 - 4.47v^4)\times 10^{-3} \nn \\
    \Gamma_{u,t}^{(2)} &=& \gamma(1.67v + 1.38v^2 - 5.46v^3 + 2.85v^4)\times 10^{-2} \nn \\
    \Gamma_{u,W}^{(2)} &=& \gamma(1.36v + 0.610v^2 - 2.90v^3 + 1.36v^4)\times 10^{-2} \nn \\
    \Gamma_{u,t}^{(3)} &=& (4.07 - 2.14v^2 + 4.76v^3 - 4.37v^4)\times 10^{-3} \nn
    \eea
    \bea
    \Gamma_{u,W}^{(3)} &=& (2.42 - 1.33v^2 + 3.14v^3 - 2.43v^4)\times 10^{-3} \nn \\
    \Gamma_{\mu,t}^{(3)} &=& (0.948v + 2.38v^2 - 4.51v^3 + 3.07v^4)\times 10^{-3} \nn \\
    \Gamma_{\mu,W}^{(3)} &=& (1.18v + 2.79v^2 - 5.31v^3 + 3.66v^4)\times 10^{-3} \nn \\
    \Gamma_{\tau,t}^{(3)} &=& (2.26v + 4.82v^2 - 9.32v^3 + 6.54v^4)\times 10^{-3} \nn \\
    \Gamma_{\tau,W}^{(3)} &=& (2.48v + 6.27v^2 - 11.9v^3 + 8.12v^4)\times 10^{-3} \nn 
\eea

Our results and differ from those of \cite{Moore:1995si}  by 
factors of $\mathcal{O}(1)$. Even taking account of the errors previously mentioned, our results are still roughly 2 times smaller. As discussed in ref.\ \cite{Kozaczuk:2015owa}, this 
discrepancy is due to the various leading log approximations made in \cite{Moore:1995si} in order to analytically evaluate  the collision integrals. Either procedure is valid to
leading accuracy, which gives an estimate of the theoretical
uncertainty associated with this approximation.
It may be worthwhile (though quite laborious) to include subleading
contributions for future studies relying upon these fluid equations.
\end{appendix}
\bibliography{ref}

\providecommand{\href}[2]{#2}\begingroup\raggedright\begin{thebibliography}{10}

\bibitem{Kajantie:1995kf}
K.~Kajantie, M.~Laine, K.~Rummukainen, and M.~E. Shaposhnikov, ``{The
  Electroweak phase transition: A Nonperturbative analysis},''
  \href{http://dx.doi.org/10.1016/0550-3213(96)00052-1}{{\em Nucl. Phys. B}
  {\bfseries 466} (1996) 189--258},
  \href{http://arxiv.org/abs/hep-lat/9510020}{{\ttfamily
  arXiv:hep-lat/9510020}}.

\bibitem{Kajantie:1996mn}
K.~Kajantie, M.~Laine, K.~Rummukainen, and M.~E. Shaposhnikov, ``{Is there a
  hot electroweak phase transition at $m_H \ge m_W$?},''
  \href{http://dx.doi.org/10.1103/PhysRevLett.77.2887}{{\em Phys. Rev. Lett.}
  {\bfseries 77} (1996) 2887--2890},
  \href{http://arxiv.org/abs/hep-ph/9605288}{{\ttfamily arXiv:hep-ph/9605288}}.

\bibitem{Trodden:1998ym}
M.~Trodden, ``{Electroweak baryogenesis},''
  \href{http://dx.doi.org/10.1103/RevModPhys.71.1463}{{\em Rev. Mod. Phys.}
  {\bfseries 71} (1999) 1463--1500},
  \href{http://arxiv.org/abs/hep-ph/9803479}{{\ttfamily arXiv:hep-ph/9803479}}.

\bibitem{Cline:2006ts}
J.~M. Cline, ``{Baryogenesis},'' in {\em {Les Houches Summer School - Session
  86: Particle Physics and Cosmology: The Fabric of Spacetime}}.
\newblock 9, 2006.
\newblock \href{http://arxiv.org/abs/hep-ph/0609145}{{\ttfamily
  arXiv:hep-ph/0609145}}.

\bibitem{Morrissey:2012db}
D.~E. Morrissey and M.~J. Ramsey-Musolf, ``{Electroweak baryogenesis},''
  \href{http://dx.doi.org/10.1088/1367-2630/14/12/125003}{{\em New J. Phys.}
  {\bfseries 14} (2012) 125003},
  \href{http://arxiv.org/abs/1206.2942}{{\ttfamily arXiv:1206.2942 [hep-ph]}}.

\bibitem{Caprini:2015zlo}
C.~Caprini {\em et~al.}, ``{Science with the space-based interferometer eLISA.
  II: Gravitational waves from cosmological phase transitions},''
  \href{http://dx.doi.org/10.1088/1475-7516/2016/04/001}{{\em JCAP} {\bfseries
  04} (2016) 001}, \href{http://arxiv.org/abs/1512.06239}{{\ttfamily
  arXiv:1512.06239 [astro-ph.CO]}}.

\bibitem{Caprini:2019egz}
C.~Caprini {\em et~al.}, ``{Detecting gravitational waves from cosmological
  phase transitions with LISA: an update},''
  \href{http://dx.doi.org/10.1088/1475-7516/2020/03/024}{{\em JCAP} {\bfseries
  03} (2020) 024}, \href{http://arxiv.org/abs/1910.13125}{{\ttfamily
  arXiv:1910.13125 [astro-ph.CO]}}.

\bibitem{Espinosa:2010hh}
J.~R. Espinosa, T.~Konstandin, J.~M. No, and G.~Servant, ``{Energy Budget of
  Cosmological First-order Phase Transitions},''
  \href{http://dx.doi.org/10.1088/1475-7516/2010/06/028}{{\em JCAP} {\bfseries
  06} (2010) 028}, \href{http://arxiv.org/abs/1004.4187}{{\ttfamily
  arXiv:1004.4187 [hep-ph]}}.

\bibitem{Konstandin:2014zta}
T.~Konstandin, G.~Nardini, and I.~Rues, ``{From Boltzmann equations to steady
  wall velocities},''
  \href{http://dx.doi.org/10.1088/1475-7516/2014/09/028}{{\em JCAP} {\bfseries
  09} (2014) 028}, \href{http://arxiv.org/abs/1407.3132}{{\ttfamily
  arXiv:1407.3132 [hep-ph]}}.

\bibitem{Huber:2013}
S.~J. Huber and M.~Sopena, ``An efficient approach to electroweak bubble
  velocities,'' \href{http://arxiv.org/abs/1302.1044}{{\ttfamily
  arXiv:1302.1044 [hep-ph]}}.

\bibitem{Megevand:2010}
A.~Mégevand and A.~D. Sánchez, ``Velocity of electroweak bubble walls,''
  \href{http://dx.doi.org/10.1016/j.nuclphysb.2009.09.019}{{\em Nuclear Physics
  B} {\bfseries 825} (2010) 151},
  \href{http://arxiv.org/abs/0908.3663}{{\ttfamily arXiv:0908.3663 [hep-ph]}}.

\bibitem{Kozaczuk:2015owa}
J.~Kozaczuk, ``{Bubble Expansion and the Viability of Singlet-Driven
  Electroweak Baryogenesis},''
  \href{http://dx.doi.org/10.1007/JHEP10(2015)135}{{\em JHEP} {\bfseries 10}
  (2015) 135}, \href{http://arxiv.org/abs/1506.04741}{{\ttfamily
  arXiv:1506.04741 [hep-ph]}}.

\bibitem{Bodeker:2009qy}
D.~Bodeker and G.~D. Moore, ``{Can electroweak bubble walls run away?},''
  \href{http://dx.doi.org/10.1088/1475-7516/2009/05/009}{{\em JCAP} {\bfseries
  05} (2009) 009}, \href{http://arxiv.org/abs/0903.4099}{{\ttfamily
  arXiv:0903.4099 [hep-ph]}}.

\bibitem{Bodeker:2017cim}
D.~Bodeker and G.~D. Moore, ``{Electroweak Bubble Wall Speed Limit},''
  \href{http://dx.doi.org/10.1088/1475-7516/2017/05/025}{{\em JCAP} {\bfseries
  05} (2017) 025}, \href{http://arxiv.org/abs/1703.08215}{{\ttfamily
  arXiv:1703.08215 [hep-ph]}}.

\bibitem{Megevand:2013hwa}
A.~Mégevand, ``{Friction forces on phase transition fronts},''
  \href{http://dx.doi.org/10.1088/1475-7516/2013/07/045}{{\em JCAP} {\bfseries
  07} (2013) 045}, \href{http://arxiv.org/abs/1303.4233}{{\ttfamily
  arXiv:1303.4233 [astro-ph.CO]}}.

\bibitem{Cline:2020jre}
J.~M. Cline and K.~Kainulainen, ``{Electroweak baryogenesis at high bubble wall
  velocities},'' \href{http://dx.doi.org/10.1103/PhysRevD.101.063525}{{\em
  Phys. Rev. D} {\bfseries 101} no.~6, (2020) 063525},
  \href{http://arxiv.org/abs/2001.00568}{{\ttfamily arXiv:2001.00568
  [hep-ph]}}.

\bibitem{Moore:1995si}
G.~D. Moore and T.~Prokopec, ``{How fast can the wall move? A Study of the
  electroweak phase transition dynamics},''
  \href{http://dx.doi.org/10.1103/PhysRevD.52.7182}{{\em Phys. Rev. D}
  {\bfseries 52} (1995) 7182--7204},
  \href{http://arxiv.org/abs/hep-ph/9506475}{{\ttfamily arXiv:hep-ph/9506475}}.

\bibitem{Moore:1995ua}
G.~D. Moore and T.~Prokopec, ``{Bubble wall velocity in a first order
  electroweak phase transition},''
  \href{http://dx.doi.org/10.1103/PhysRevLett.75.777}{{\em Phys. Rev. Lett.}
  {\bfseries 75} (1995) 777--780},
  \href{http://arxiv.org/abs/hep-ph/9503296}{{\ttfamily arXiv:hep-ph/9503296}}.

\bibitem{John:2000zq}
P.~John and M.~Schmidt, ``{Do stops slow down electroweak bubble walls?},''
  \href{http://dx.doi.org/10.1016/S0550-3213(00)00768-9}{{\em Nucl. Phys. B}
  {\bfseries 598} (2001) 291--305},
  \href{http://arxiv.org/abs/hep-ph/0002050}{{\ttfamily arXiv:hep-ph/0002050}}.
  [Erratum: Nucl.Phys.B 648, 449--452 (2003)].

\bibitem{Huber:2011aa}
S.~J. Huber and M.~Sopena, ``{The bubble wall velocity in the minimal
  supersymmetric light stop scenario},''
  \href{http://dx.doi.org/10.1103/PhysRevD.85.103507}{{\em Phys. Rev. D}
  {\bfseries 85} (2012) 103507},
  \href{http://arxiv.org/abs/1112.1888}{{\ttfamily arXiv:1112.1888 [hep-ph]}}.

\bibitem{Dorsch:2018pat}
G.~C. Dorsch, S.~J. Huber, and T.~Konstandin, ``{Bubble wall velocities in the
  Standard Model and beyond},''
  \href{http://dx.doi.org/10.1088/1475-7516/2018/12/034}{{\em JCAP} {\bfseries
  12} (2018) 034}, \href{http://arxiv.org/abs/1809.04907}{{\ttfamily
  arXiv:1809.04907 [hep-ph]}}.

\bibitem{Joyce:1994zt}
M.~Joyce, T.~Prokopec, and N.~Turok, ``{Nonlocal electroweak baryogenesis. Part
  2: The Classical regime},''
  \href{http://dx.doi.org/10.1103/PhysRevD.53.2958}{{\em Phys. Rev. D}
  {\bfseries 53} (1996) 2958--2980},
  \href{http://arxiv.org/abs/hep-ph/9410282}{{\ttfamily arXiv:hep-ph/9410282}}.

\bibitem{Note1}
In the fluid frame these are simply $1$, $p_z$ and $E$.

\bibitem{Note2}
Strictly speaking, this method only works when the $z$-dependence of
  $A_v^{-1}\Gamma $ can be ignored on either side of the wall, but the same
  conclusion is borne out by a full numerical solution.

\bibitem{Cline:2000nw}
J.~M. Cline, M.~Joyce, and K.~Kainulainen, ``{Supersymmetric electroweak
  baryogenesis},'' \href{http://dx.doi.org/10.1088/1126-6708/2000/07/018}{{\em
  JHEP} {\bfseries 07} (2000) 018},
  \href{http://arxiv.org/abs/hep-ph/0006119}{{\ttfamily arXiv:hep-ph/0006119}}.

\bibitem{Fromme:2006cm}
L.~Fromme, S.~J. Huber, and M.~Seniuch, ``{Baryogenesis in the two-Higgs
  doublet model},'' \href{http://dx.doi.org/10.1088/1126-6708/2006/11/038}{{\em
  JHEP} {\bfseries 11} (2006) 038},
  \href{http://arxiv.org/abs/hep-ph/0605242}{{\ttfamily arXiv:hep-ph/0605242}}.

\bibitem{Fromme:2006wx}
L.~Fromme and S.~J. Huber, ``{Top transport in electroweak baryogenesis},''
  \href{http://dx.doi.org/10.1088/1126-6708/2007/03/049}{{\em JHEP} {\bfseries
  03} (2007) 049}, \href{http://arxiv.org/abs/hep-ph/0604159}{{\ttfamily
  arXiv:hep-ph/0604159}}.

\bibitem{Arnold:2000}
P.~Arnold, G.~D. Moore, and L.~G. Yaffe, ``{Transport coefficients in high
  temperature gauge theories (I): leading-log results},''
  \href{http://dx.doi.org/10.1088/1126-6708/2000/11/001}{{\em JHEP} {\bfseries
  0011} (2000) 001}, \href{http://arxiv.org/abs/hep-ph/0010177}{{\ttfamily
  arXiv:hep-ph/0010177 [hep-ph]}}.

\bibitem{Note3}
$L/\gamma $ is the wall thickness as measured in the plasma frame. Fixing
  $LT/\gamma $ rather than $LT$ makes it easier to see that the diffusion tails
  in front of the wall disappear as $v\to 1$.

\bibitem{Cline:2021iff}
J.~M. Cline, A.~Friedlander, D.-M. He, K.~Kainulainen, B.~Laurent, and
  D.~Tucker-Smith, ``{Baryogenesis and gravity waves from a UV-completed
  electroweak phase transition},''
  \href{http://arxiv.org/abs/2102.12490}{{\ttfamily arXiv:2102.12490
  [hep-ph]}}.

\bibitem{Weldon:1982bn}
H.~A. Weldon, ``{Effective Fermion Masses of Order gT in High Temperature Gauge
  Theories with Exact Chiral Invariance},''
  \href{http://dx.doi.org/10.1103/PhysRevD.26.2789}{{\em Phys. Rev. D}
  {\bfseries 26} (1982) 2789}.

\bibitem{Note4}
As pointed out in ref.\ \cite {Arnold:2000}, there were some errors in the
  expressions of the scattering amplitudes in \cite {Moore:1995si}. They failed
  to include a $1/2$ symmetry factor in the amplitude for $\protect \bar {t}t
  \to gg$ and made some algebraic errors in $tq \to tq$ and $Wf \to Wf$.

\bibitem{Arnold:2003}
P.~Arnold, G.~D. Moore, and L.~G. Yaffe, ``Transport coefficients in high
  temperature gauge theories: (ii) beyond leading log,''
  \href{http://dx.doi.org/10.1088/1126-6708/2003/05/051}{{\em Journal of High
  Energy Physics} {\bfseries 0305} (2003) 051},
  \href{http://arxiv.org/abs/hep-ph/0302165}{{\ttfamily arXiv:hep-ph/0302165
  [hep-ph]}}.

\bibitem{Moore:2001}
G.~D. Moore, ``Transport coefficients in large nf gauge theory: testing hard
  thermal loops,'' \href{http://dx.doi.org/10.1088/1126-6708/2001/05/039}{{\em
  Journal of High Energy Physics} {\bfseries 0105} (2001) 039},
  \href{http://arxiv.org/abs/hep-ph/0104121}{{\ttfamily arXiv:hep-ph/0104121
  [hep-ph]}}.

\end{thebibliography}\endgroup
\bibliographystyle{utphys}

\end{document}